\pdfoutput=1 
\documentclass[fleqn,useAMS,usenatbib,usedcolumn]{mnras}

\usepackage{float}

\usepackage[british]{babel}             
\usepackage{newtxtext}                  
\usepackage[slantedGreek]{newtxmath}    

\let\la=\lesssim 

\usepackage{aasmacros}
\usepackage{amssymb,amsmath} 
\usepackage[pdftex]{graphicx}
\usepackage{url}
\usepackage{times} 
\usepackage{xcolor}

\usepackage[T1]{fontenc}
\usepackage{aecompl}

\hypersetup{pdfauthor={A. P. Cooper},
               pdftitle={Comparing semi-analytic particle tagging and hydrodynamical simulations of the Milky Way's stellar halo},
               pdfkeywords={methods: numerical, galaxies: dwarf, galaxies: haloes, galaxies: structure},
               bookmarksnumbered=true}

\newcommand{\ie}{i.e.\ }
\newcommand{\eg}{e.g.\ }

\newcommand{\Msol}{\,\mathrm{M_{\sun}}}
\newcommand{\kpc}{\,\mathrm{kpc}}

\newcommand{\CC}[1]{#1}
\newcommand{\checkme}[1]{\textcolor{black}{#1}}
\newcommand{\BB}[1]{\textcolor{black}{#1}}

\newcommand{\galform}{\textsc{galform}}
\newcommand{\gasoline}{\textsc{gasoline}}
\newcommand{\sphsim}{AqC-SPH}
\newcommand{\dmsim}{AqC-DM}
\newcommand{\stings}{\textsc{stings}}
\newcommand{\subfind}{\textsc{subfind}}

\DeclareRobustCommand{\fmb}{f_{\mathrm{mb}}}
\DeclareRobustCommand{\fnz}{f_{\mathrm{90}}}
\DeclareRobustCommand{\theo}{\citet{Le-Bret:2015aa}}
\DeclareRobustCommand{\ins}{\textit{in situ}}

\title[Comparing particle tagging to SPH]{Comparing semi-analytic particle tagging and
hydrodynamical simulations of the Milky Way's stellar halo}

\author[Cooper et al.]{Andrew P. Cooper$^{1}$\thanks{E-mail:
a.p.cooper@durham.ac.uk}, {Shaun Cole$^{1}$}, {Carlos S. Frenk$^{1}$}, {Theo Le Bret$^{2,3}$} and {Andrew Pontzen$^{2}$}\\
$^{1}$Institute for Computational
Cosmology, Department of Physics, University of Durham, South Road, Durham, DH1
3LE, UK\\
$^{2}$Department of Physics and Astronomy, University College London, Gower Street, London, WC1E 6BT, UK\\
$^{3}$Rudolf Peierls Centre for Theoretical Physics, University of Oxford, Oxford, OX1 3NP, UK\\
}

\volume{{\rm in press}}  

\begin{document}

\date{Accepted 2017 April 14. Received 2017 March 17; in original form 2016 November 10}

\pagerange{\pageref{firstpage}--\pageref{lastpage}} \pubyear{2017}

\maketitle

\label{firstpage}

\begin{abstract} 

  Particle tagging is an efficient\BB{, but approximate,} technique for using
  cosmological $N$-body simulations to model the phase-space evolution of the
  stellar populations predicted\BB{, for example,} by a semi-analytic model of
  galaxy formation. We test the technique developed by Cooper et al. (which we
  call \stings{} here) by comparing particle tags with stars in a smooth
  particle hydrodynamic (SPH) simulation. We focus on the spherically averaged
  density profile of stars accreted from satellite galaxies in a Milky Way
  (MW)-like system.  The stellar profile in the SPH simulation can be recovered
  accurately by tagging dark matter (DM) particles in the same simulation
  according to a prescription based on the rank order of particle binding
  energy.  Applying the same prescription to an $N$-body version of this
  simulation produces a density profile differing from that of the SPH
  simulation by $\lesssim10$~per cent on average between 1 and 200~kpc. This
  confirms that particle tagging can provide a faithful and robust
  approximation to a self-consistent hydrodynamical simulation in this regime
  (in contradiction to previous claims in the literature). We find only one
  systematic effect, likely due to the collisionless approximation, namely that
  massive satellites in the SPH simulation are disrupted somewhat earlier than
  their collisionless counterparts.  In most cases this makes remarkably little
  difference to the spherically averaged distribution of their stellar debris.
  We conclude that, for galaxy formation models that do not predict strong
  baryonic effects on the present-day DM distribution of MW-like galaxies or
  their satellites, differences in stellar halo predictions associated with the
  treatment of star formation and feedback are much more important than those
  associated with the dynamical limitations of collisionless particle tagging.

\end{abstract}

\begin{keywords}
  methods: numerical --  galaxies: dwarf  -- galaxies: haloes -- galaxies: structure 
\end{keywords}

\section{Introduction}

A number of studies have used so-called particle tagging techniques to predict
the distribution and kinematics of Milky Way (MW) halo stars accreted from tidally
disrupted dwarf satellite galaxies \citep{Bullock01_BKW,Bullock05, DeLucia08,
Tumlinson:2010aa, Cooper:2010aa, Libeskind11, Rashkov12}. These techniques attempt
to model both stars and dark matter (DM) with a single collisionless particle
species in a cosmological $N$-body simulation \checkme{by `painting' subsets of the
particles with stellar mass, according to a weighting function, without
changing the mass of the particle used in the gravitational calculation}. This
is intended as an approximation to the more self-consistent approach of
hydrodynamical simulations, in which a separate species of collisionless `star
particles' is inserted into the calculation to replace gas particles that
become sufficiently cold and dense.  This replacement is usually done according
to a `subgrid' model of star formation describing the state of the
interstellar medium represented by the gas particle
\citep[e.g.][]{Schaye:2015aa}. In particle tagging models, all the baryonic
physics of galaxy formation (including dissipative cooling, star formation and
feedback of mass and energy to the interstellar medium) are modelled
semi-analytically\footnote{An alternative approach ignores the physics of
galaxy formation and instead assigns stellar mass directly to DM haloes using
theoretical or empirical scaling relations
\citep[e.g.][]{Bullock05,Rashkov12,Laporte13b}.} on the scale of DM haloes
\citep{DeLucia08, Cooper:2010aa}. The assignment of stellar mass to DM is usually
\BB{expressed} as a function of the binding energy of the DM particles, in
order to account for the prior dissipation of energy by the star-forming gas
\citep{Bullock01,Bullock05,Penarrubia08}. 

The most significant differences between tagging schemes concern those two
aspects of the approach -- how the star formation histories (SFHs) of DM haloes are
computed, and the algorithm used to associate stellar mass with specific
particles in the $N$-body simulation. In \citet[][hereafter C10]{Cooper:2010aa} we
described a technique in which a semi-analytic model (\galform{}, in our case)
is used to predict SFHs, \BB{and the tagging operation
is carried out for every star-forming halo at every snapshot (so-called `live'
tagging; the common alternative is to tag DM only once per satellite halo, at
the time when it crosses the virial radius of the `main' halo).} In each halo,
$N$-body particles are ranked in order of binding energy and the stellar mass
to be assigned is distributed equally among a fixed fraction
($f_{\mathrm{mb}}$) of the most bound (in \citealt{Cooper:2010aa}, $\fmb=1$~per
cent). This technique was applied to MW-like galaxies by C10 and later
extended to more massive galaxies by \citet[][C13]{Cooper:2013aa} and
\citet{Cooper:2015ab}, who give additional details of the method.  To
distinguish the scheme set out in those three papers from other particle
tagging schemes in the literature, we refer to it hereafter as \stings{}
(\textit{Stellar Tags In $N$-body Galaxy Simulations}).

The advantage of particle tagging over hydrodynamical methods is that the
evolving phase-space distribution of the stellar component can be followed at
much higher resolution and at much lower computational cost. This in turn allows
phenomena that arise from the dynamics of hierarchical assembly, including
stellar haloes and the scaling relations of elliptical galaxies, to be explored
with a much wider range of galaxy formation models. In that sense, the
particle tagging approximation can be thought of as an extension of the
semi-analytic approach to modelling galaxy formation \citep[a comprehensive
overview of which is given by][]{Lacey:2016aa}.

It is very clear that particle tagging is only an approximate technique,
because the contribution of baryons to the gravitational potential is not
treated self-consistently in the dynamical part of the calculation. In that
calculation, each $N$-body particle includes a baryonic mass equal to the
universal fraction, regardless of how much stellar mass the tagging procedure
associates with it and regardless of the inflow and outflow of gas assumed in
the semi-analytic component of the model. The aim of this paper is to use
hydrodynamical simulations as a benchmark to test how this approximation
affects \BB{some of the most basic} predictions that have been made with
particle tagging methods. We use \stings{} as the basis for our comparison, but
our results are relevant to the particle tagging technique more generally. We
present our results in the context of the MW and its stellar halo
because this is the regime in which particle tagging has been applied most
often.  In that context, the most important astrophysical processes that
typical $N$-body particle tagging schemes neglect are as follows:

\begin{enumerate}
    
 \item The \textit{internal structure} of the DM haloes of satellite
   galaxies can be altered by the inflow and outflow of baryons. The rapid
   dissipative condensation of gas within haloes can increase the central
   density of the DM (`cusp formation'), and the rapid gas expulsion of
   gas by supernova feedback may have the opposite effect (`core formation';
   \eg   \citealt{Navarro:1996aa, Pontzen:2012aa, Nipoti:2015aa} and references
   therein).  Both contraction and expansion of the potential can affect the
   kinematics of stars and the resulting rate of mass loss through tidal forces
   \citep{Penarrubia:2010aa,Errani:2015aa,Read:2016aa}.  This implies that  the method should
   be restricted to satellite galaxies with high mass-to-light ratios, although
   tagging has also been used to make predictions for the structure of
   stellar haloes in more massive galaxies \citep[\eg][]{Cooper:2013aa}.

\item The \textit{orbital evolution} of a particular satellite may be different
  in simulations from the same initial conditions with and without
  hydrodynamics, because the host halo potential can also change shape and
  concentration in response to the motion of baryons \citep[\eg][]{Abadi:2010aa,
  Binney:2015aa}.  The growth of a massive stellar disc could make the initially
  triaxial inner regions of the host halo more oblate or spherical.  Differences
  in the rate of mass loss due to these changes or other effects (such as ram
  pressure stripping of gas; \eg  \citealt{Arraki:2014aa}) could exacerbate
  initially small divergences in satellite orbits. 

\item Changes in the host potential will also affect \textit{strong
  gravitational interactions} involving satellites and associated dynamical
  heating and disruptive effects, particularly for satellites with orbits that
  pass through the centre of the host ($r<\sim20$~kpc). Disc shocking
  \citep{Spitzer:1973aa} is an example of this kind of interaction
  \citep[\eg][]{DOnghia:2010aa}. The consequences will depend on the strength
  and extent of the perturbation represented by the disc, its evolution with
  time, and the number of halo-progenitor satellites that pass through the
  region concerned.

\item In hydrodynamic simulations, stars can form on phase-space trajectories
  that are not well-sampled in an equivalent collisionless $N$-body simulation,
  most notably those in centrifugally supported discs.
  Likewise, subject to the hydrodynamic scheme and sub-grid star formation
  prescription, gas particles stripped from infalling satellites may spawn stars directly
  on halo-like orbits \citep{Cooper:2015aa}.  Consequently, applications of the
  method have mostly been restricted to the accreted component of galactic
  stellar haloes, as opposed to their possible \ins{} components. 

\end{enumerate}

Earlier `comparative' tests along similar lines to ours have been carried out
by \citet{Libeskind11} and \citet{Bailin:2014aa}. A more detailed discussion of
these earlier studies is given in Section~\ref{sec:discussion}. In summary, the
interpretation of this previous work is complicated by (i) the introduction of
more complex tagging schemes \citep{Libeskind11} or simplified schemes which
omit important features of those commonly used in the literature
\citep{Bailin:2014aa}; (ii) the use of hydrodynamic galaxy formation models
that do not match well to observations in the regime under study; (iii) the use
of complex statistics, such as the clustering of the projected stellar density
\citep{Bailin:2014aa} as a basis for comparison; and (iv) the lack of a
sufficiently clear distinction between uncertainties that are directly
associated with the dynamical approximations involved in particle tagging and
uncertainties associated with other differences in the models being compared.
The comparison we present here addresses all these points. Point (iii) and (iv)
are particularly important because the extent to which particle tagging `works'
as a proxy for a given hydrodynamical model depends on whether the
approximations of the tagging scheme are justified for the conditions under
which stars form in that particular model. When interpreting tests of particle
tagging based on such comparisons, it should be borne in mind that different
smooth particle hydrodynamics (SPH) simulations and semi-analytic models
currently make very different predictions for when, where and in what quantity
stars form, particularly in dwarf galaxies. In addition to the discussion
below, we refer the reader to \theo{}, for another perspective on this issue
and its implications for the dynamical evolution of sets of tagged particles in
hydrodynamical and collisionless simulations.  Finally, we note that
\citet{Dooley:2016aa}, in their study of stellar haloes in self-interacting DM
models, also compare the predictions of a tagging prescription similar to ours
against an SPH simulation from the same initial conditions. The stellar masses
in their tagging model were obtained from an abundance matching argument,
rather than a forward model of star formation; the only control for differences
in stellar mass and SFH between their collisionless and SPH
realization was a post-hoc renormalization of total stellar mass by an order of
magnitude.  Nevertheless, \citet{Dooley:2016aa} found close agreement in
stellar mass density between 50 and 200~kpc.

We proceed as follows. We concentrate on the spherically averaged density
profile of the stellar halo, a particularly straightforward and relevant
prediction which features prominently in previous work using particle tagging
\citep{Bullock05, DeLucia08, Tumlinson:2010aa, Cooper:2010aa, Libeskind11}. We
examine high resolution cosmological SPH simulations of galaxies similar to the
MW (Section~\ref{sec:simulations}). 
In Section~\ref{sec:basiccomparison}, we first use the DM distribution in our
SPH simulations together with the SFH of each halo in the
\textit{same} simulation to produce a particle tagging approximation for the
distribution of stellar mass, which we compare to the original SPH star
particles (Section~\ref{sec:sphtag}).  We then repeat this exercise with a separate
$N$-body simulation that starts from initial conditions identical to those of
our SPH simulation (Section~\ref{sec:dmtag}). These are our main results.
Section~\ref{sec:details} examines in detail the origin of the (small)
discrepancies we find between our SPH simulation and \stings{} applied to the
$N$-body version. We examine the choice of $\fmb$, the single free parameter
in the C10 implementation of \stings{} (Section 5). In Section 6 we discuss our
findings in the context of previous work on particle tagging. We summarize our
results in Section 7.  In Appendices A and B, we present examples that
illustrate why (in the `comparative' approach used here and in previous work on
this topic), it is important to distinguish systematic and stochastic
discrepancies that arise from modelling the collisional dynamics of baryons
explicitly from less relevant effects that arise from the use of different
models for star formation.  Appendix C revisits how \citet{Cooper:2010aa} used size
distribution of satellite galaxies as a constraint on particle tagging models
in light of the results here, and Appendix D discusses numerical convergence.

This paper is about comparing particle tagging and SPH simulations, rather than
comparing either of these methods to observational data on stellar haloes in
detail.  Readers who are more interested in the `bottom line' performance of
particle tagging schemes in the context of the MW's stellar halo than in
the technicalities of the method might, therefore, prefer to examine the first
two figures and related text in Section~\ref{sec:basiccomparison} and then skip
ahead to the comparison with earlier work in Section 6 and the summary of our
findings in Section~7.

\section{Simulations}
\label{sec:simulations}

Our hydrodynamic simulation\footnote{Although we only discuss tests based on
\sphsim{} in detail here, we refer the reader to \theo{}, who, in the context
of particle tagging, compare this simulation to others based on a different
code, \gasoline{,} with alternative subgrid physical recipes.}, which we refer
to as \sphsim{}, is described by \citet{Parry12} and \citet{Cooper:2015aa}. It
uses initial conditions from the Aquarius project \citep{Springel08},
specifically those of halo Aq-C at resolution level 4, as the basis for
resimulation with an upgraded version of the SPH scheme described by
\citet{Okamoto:2010aa}. Particle masses are $2.6\times10^{5} \Msol$ for DM and
$5.8\times10^{4}\Msol$ for gas (assuming the Hubble parameter $h=0.73$). The
Plummer-equivalent softening length is $\epsilon_{\mathrm{phys}}=257$~pc. 128
snapshots of the simulation were stored, spaced evenly by 155~Myr at redshifts
$z<2.58$ and by shorter intervals at higher redshift. The virial mass of the
MW analogue is $1.8\times10^{12}\Msol$, towards the upper end of
constraints on the most likely MW halo mass from recent measurements
\citep[see for example the compilation of results in fig. 1 
of][]{Wang:2015aa}. A stable baryonic disc forms at $z\sim2.5$ and persists to
$z=0$ \citep[see][]{Scannapieco:2012aa}.  Excluding self-bound satellites, the
total stellar mass bound to the main halo at $z=0$ (comprising the disc and
spheroid of the MW analogue) is $4.1\times10^{10}\Msol$. We also make
use of a DM-only version of this simulation (halo Aq-C-4 of
\citealt{Springel08}) with the same initial density perturbation phases and
comparable resolution, which we refer to as \dmsim{}.  \citet{Zhu:2016aa} have
recently examined the properties of satellites simulated from these same
initial conditions with a moving-mesh hydrodynamical scheme.

\section{Particle Tagging}
\label{sec:basiccomparison}

\subsection{Tagging in an SPH simulation}
\label{sec:sphtag}

\begin{figure}
  \includegraphics[width=84mm, trim=0.0cm 0cm 0.0cm 0cm,clip=True]{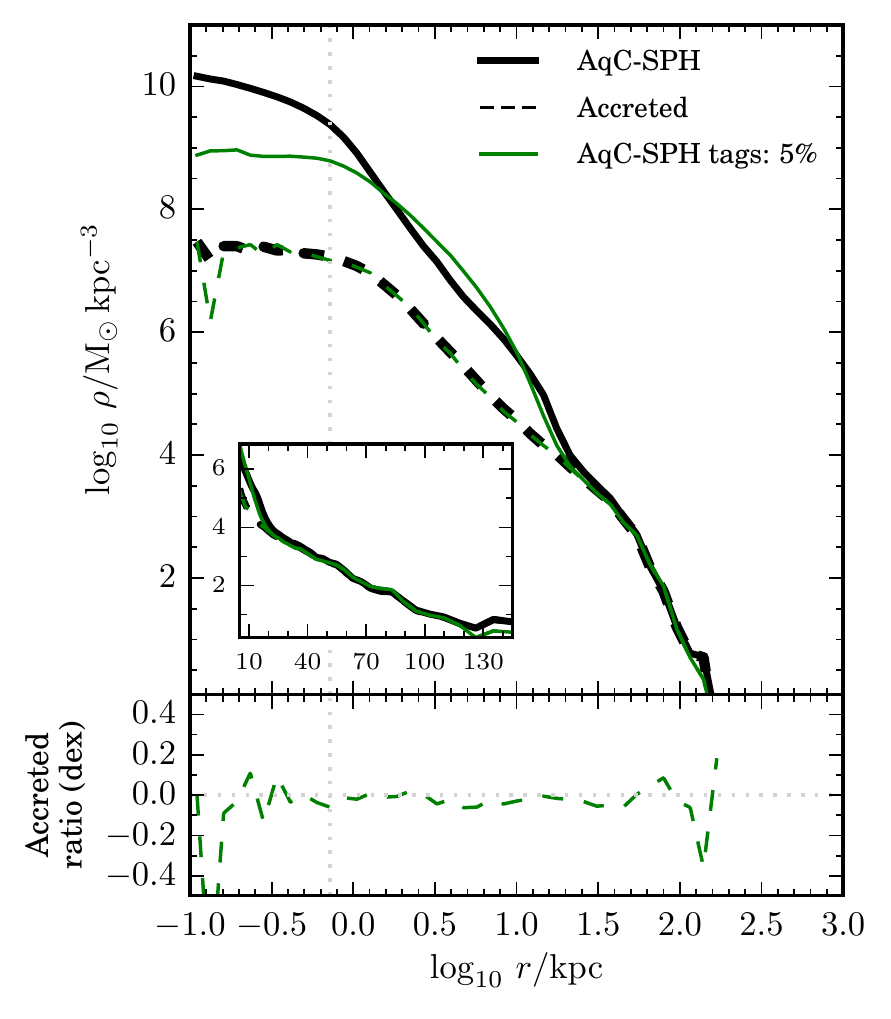}

  \caption{Top: spherically averaged mass density profile of all star particles
  bound to the main halo in \sphsim{} (black) compared with that obtained by
  tagging DM particles in \sphsim{} based on the SPH SFH (green), using
  $\fmb=5$ per cent. Dashed lines of the same colours show accreted stars only.
  The inset shows the same curves on a linear radial scale from 5 to 150 kpc.
  Bottom: logarithm of the ratio between the SPH and tagged-particle density
  profiles for accreted stars. This figure demonstrates that the distribution
  of star particles and tagged particles in the same SPH simulation agree well,
  particularly for the accreted component, even for a tagging scheme based only
  on rank order of particle binding energy.}

\label{fig:c4_main_sph_tags}
\end{figure}

The first question we ask is how well DM particle tagging works \textit{within}
our SPH simulation. This is obviously not how particle tagging is applied in
practice but it provides a  benchmark for interpreting the differences that
arise when we tag particles in collisionless simulations.

The mass, formation time and phase-space trajectory of each star particle in
\sphsim{} are known precisely. We extract SFHs by building
a merger tree and assigning each of these star particles to the halo or subhalo
to which it is bound at the first snapshot following its formation. We then use
these self-consistent SFHs as the `input' for tagging of DM
particles in the same \sphsim{} haloes, following the \stings{} scheme outlined
in C10. In this experiment, tagged DM particles and the `original' star
particles experience the same orbital evolution and tidal field, as in
\citet{Bailin:2014aa} and \citet{Le-Bret:2015aa}. \BB{This simple experiment
allows us to study directly the effects of the differences in the initial
phase-space distribution of star particles and tagged DM particles that 
result from the approximations inherent in the tagging procedure.}

Fig.~\ref{fig:c4_main_sph_tags} compares the stellar mass density profile of
the MW analogue\footnote{All star particles bound to the main halo,
comprising the disc, spheroid and halo of the main galaxy and excluding
satellites.} in \sphsim{} at $z=0$ with the analogous result obtained by
tagging DM particles according to the `fixed-fraction' \stings{} scheme, using
$\fmb=5$ per cent\footnote{Since the point of this exercise is to compare the
two simulation techniques rather than to interpret observations, the
appropriate value of $\fmb$ is that which best approximates the behaviour of
the subgrid star formation model in \sphsim{} with regard to the distribution
of stellar binding energies after dissipative collapse.} and the
`self-consistent' SPH SFHs. These curves are the same as
those in the lower right panel of fig.~2 in \citealt{Le-Bret:2015aa}. Note
that here both the star particle and tagged particle profiles \textit{include}
the \ins{} stellar component (the figures in C10 did not show this component).
The agreement is reasonably close, with discrepancies of no more than an order
of magnitude at any radius over a density range covering ten orders of
magnitude \checkme{and little or no discernible systematic offset}.
\checkme{Physical features in the profile, such as the in-situ to accreted
transition ($r\gtrsim10$~kpc) and `breaks' in the density of accreted stars,
are much more significant than these discrepancies. The most obvious
differences between star particles and tagged DM particles are seen in the
inner 10~kpc.  These are the result of differences in the density of \ins{}
stars, since accreted stars contribute very little mass in these regions. } 

If we only consider the accreted stellar component (as in C10 and most other
applications of particle tagging) the agreement between star particles and tags
is much closer \checkme{at all radii (dashed lines), well below $0.1$~dex for
$r<100$.  This implies that the dynamical differences between the two
techniques will not dominate the uncertainty in typical comparisons to real
data, for example on the shape and amplitude of accreted stellar halo density
profiles or their moments, such as total mass and half-mass radius. The
observational errors on these quantities are of a similar order ($\sim
0.5$~dex; e.g. the density of the MW stellar halo in the Solar neighbourhood;
\citealt{McKee:2015aa}) and the system-to-system scatter likely greater (for
example the density of the stellar halo of MW-like galaxies at 30~kpc has a
scatter of $>1$~dex; \citealt{Cooper:2013aa}).}

In Section~\ref{sec:nearest} we will show that the mismatch between
the spherically averaged density profiles of \ins{} star particles and their
corresponding tags in Fig.~\ref{fig:c4_main_sph_tags} arises because a single,
universal value of $\fmb$ cannot adequately represent the complex energy
distribution of star-forming gas particles in the inner regions of our MW
analogue. An alternative explanation for this mismatch might be the
three-dimensional (3D) shape of the \ins{} component, which is highly oblate in
our SPH simulation. Most star particles belong to a rotationally supported
disc, which obviously cannot be reproduced by DM particles selected on the
basis of energy alone (even in this case, where the DM particles also feel the
potential generated by the stellar disc).  However, Section~\ref{sec:nearest}
demonstrates that the difference in 3D shape is \textit{not} responsible for
the majority of the discrepancy seen in Fig.~\ref{fig:c4_main_sph_tags}.

\subsection{Tagging in a collisionless simulation}
\label{sec:dmtag}

\begin{figure}
  \includegraphics[width=84mm, trim=0.2cm 0.2cm 0.1cm 0.2cm,clip=True]{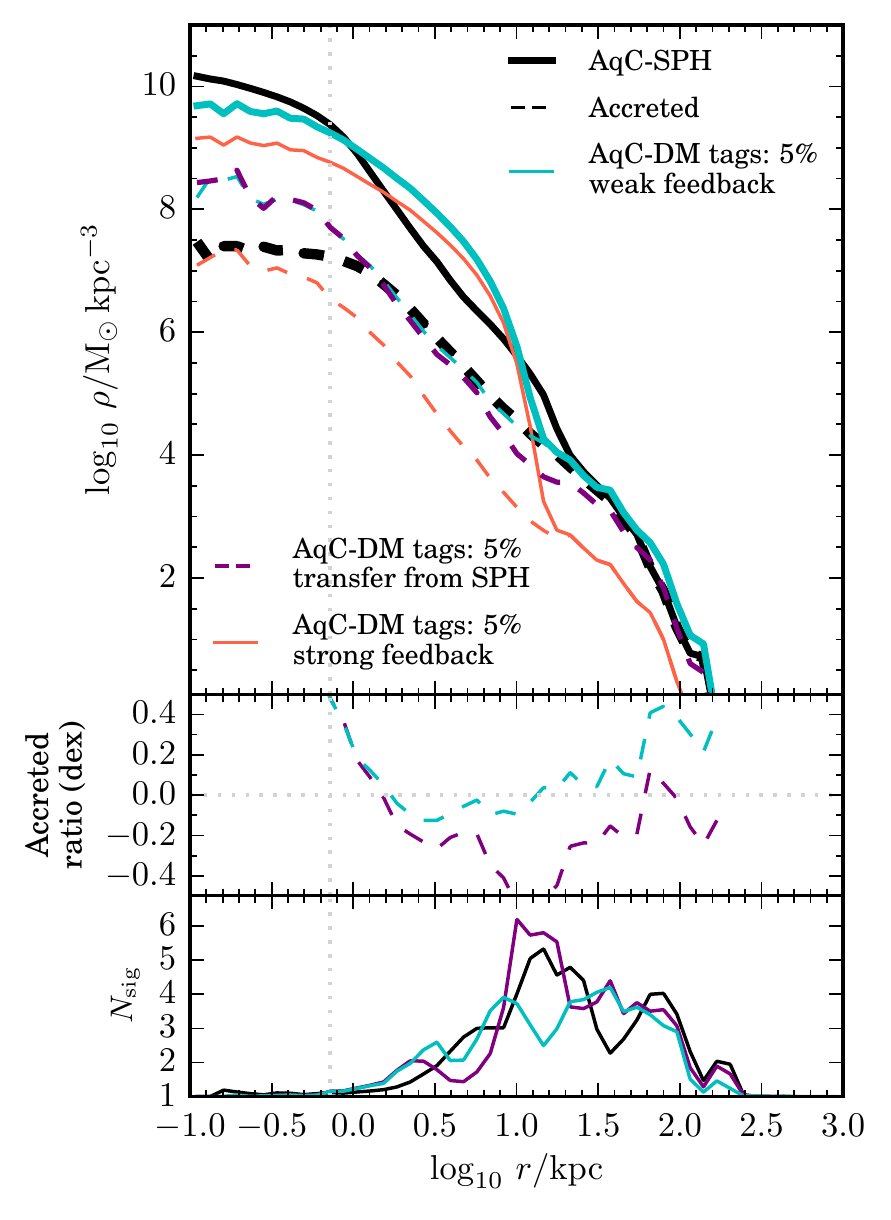}

  \caption{Top: spherically averaged mass density profile of all star particles
  bound to the main halo in \sphsim{} (black) compared with those obtained by
  tagging DM particles in a collisionless simulation with the same
  initial conditions, \dmsim{}, based on semi-analytic SFHs
  predicted by the \citet{Bower:2006aa} \galform{} model (which has `strong'
  feedback; orange) and a variant of this model with `weak' feedback (cyan).
  Also shown (purple) is the profile resulting from tagging based on a direct
  transfer of the SFHs from the SPH simulation to the
  collisionless simulation for the 10 most massive halo progenitors only.
  Dashed lines show only the accreted stellar component. All tagging results
  use $\fmb=5$ per cent.  Middle: ratio of accreted density profiles as in
  Fig.~\ref{fig:c4_main_sph_tags}. Bottom: the `diversity' of the stellar halo;
  lines show $N_{\mathrm{sig}}(r)=[\Sigma_{i} m_{i}(r)]^{2} /
  \Sigma_{i}[m_{i}(r)^{2}]$, an estimate of the number of progenitors
  contributing a significant fraction of accreted debris at each radius.  The
  good agreement seen in Fig.~\ref{fig:c4_main_sph_tags} holds for this
  application to a collisionless simulation, which reflects how particle tagging is
  most commonly applied in practice. The additional discrepancies are dominated
  by differences in star formation modelling rather than particle tagging (see
  the text).}

\label{fig:c4_main_galform}
\end{figure}

The SPH experiment above was designed to take the effects of baryons on the
galactic potential out of the comparison, to demonstrate that the phase-space
evolution of stellar populations in our simulation is then well approximated by
the DM particles we select as tags. In practice (\eg in C10), particle tagging
is used to model the phase-space distribution of stars in collisionless
simulations, which do not include baryonic effects on gravitational dynamics.
We now proceed to a more general comparison between \sphsim{} and \stings{}
applied to \dmsim{}, a `DM-only' simulation from the same initial conditions.
The mass of particles in \dmsim{} includes the universal fraction of baryonic
mass and is therefore 
\BB{larger than that of the DM particles in \sphsim{} by a factor
$\Omega_{0}/(\Omega_{0} - \Omega_{\mathrm{baryon}})$, where
$\Omega_{\mathrm{baryon}}=0.045$}.  

\subsubsection{`Transplanting' of SFHs from Aq-SPH}
\label{sec:direct_transfer}

For the experiment in this section, we need to assign SFHs to haloes in
\dmsim{}.  This would be straightforward if the SFH of every halo could be
`transplanted' on a one-to-one basis from \sphsim{}. However, the dynamical
histories of corresponding haloes in the two simulations sometimes diverge,
which means that any transplanting is unavoidably approximate, particularly in
the highly nonlinear regime of tidally disrupting satellite galaxies. Some of
this divergence is the direct result of baryonic physics
\citep[e.g.][]{Sawala:2015aa}; some is stochastic or an indirect consequence of
other physical changes; and some is simply the result of ambiguities in the
numerical methods we use to identify haloes and subhaloes and link them between
snapshots. Transplanting SFHs therefore requires a degree of care that makes it
impractical to take this approach for every halo in the simulation.

For these reasons, we carry out a careful individual analysis for only the 10 progenitor
haloes that make the most significant contributions of mass to the stellar
halo in \sphsim{}. We find that nine of these progenitors each contribute more
than 1 per cent of the total mass of stars accreted by our MW analogue;
the most massive contributes 30 per cent. 
For simplicity, we control for the fact that some
of these satellites accrete a (small) fraction of their stars from their own
hierarchical progenitors by considering only star formation in the `main
branch' of the merger tree of each satellite when transplanting their SFHs.

Fig.~\ref{fig:c4_main_galform} compares the density profile of tagged particles
in \dmsim{} obtained with this transplanting operation (dashed purple line)
against the profiles of star particles in \sphsim{} (the figure also shows
profiles of tags in \dmsim{} based on \galform{} SFHs,
which are described in the following subsection). We only compare the accreted
stellar component because it makes little sense to tag stars formed \ins{} with
a procedure like this (more details on this point are given below). The
`transplanted' profile agrees well with the SPH result (over five orders of
magnitude in density and two in radius above the gravitational softening scale)
except for a $\sim 30$~per cent discrepancy in the region $10<r<30 \,
\mathrm{kpc}$. In this region the contributions to the stellar halo from
different progenitors are most equal. The statistic
$N_{\mathrm{sig}}(r)=[\Sigma_{i} m_{i}(r)]^{2} / \Sigma_{i}[m_{i}(r)^{2}]$,
where $m_{i}(r)$ is the stellar mass contributed by the $i$th progenitor at
radius $r$, implies that approximately five progenitors each contribute $1/5$ of
the stellar mass in this region, in both \sphsim{} and the transplant-tagged
version of \dmsim{}.  Relatively small changes to the balance between the
different contributions therefore have a particularly notable effect in this
region. 

We find no single progenitor dominates the differences we see in
Fig.~\ref{fig:c4_main_galform}. Instead we find that satellites in \sphsim{}
tend to be disrupted (\ie no longer identified as self-bound systems by
\subfind{}) several Gyr earlier than their counterparts in \dmsim{}.
However, this has a surprisingly small effect on the spherically averaged
distribution of their debris in most cases. The characteristic apocentres of
the debris are significantly smaller in \sphsim{} in only two case. A third
case, the most massive of all the halo contributors, exhibits the same effect
but to a lesser extent. We find that the stellar mass density profiles of the
individual progenitors at the time of their infall to the main halo are very
similar in the two simulations. The differences in their debris distributions
must therefore arise from how and when the individual satellites are disrupted.
These effects are most notable in the inner region of the halo, where we
would expect baryons in the central galaxy to create the most significant
differences in the potential. The above results concerning the details of how
and why individual satellites differ are presented in
Appendix~\ref{sec:why_do_individuals_differ}.

\BB{We conclude that neither the self-consistent treatment of baryonic physics
in \sphsim{} nor stochastic differences in the orbits of subhaloes between
\sphsim{} and \dmsim{} give rise to discrepancies with particle tagging that
would significantly change our interpretation of predictions for the
spherically averaged density profile of the accreted stellar halo.} 

\subsubsection{Semi-analytic SFHs}
\label{sec:galform_tagging}

The final, most general step in our comparison is to use the \galform{}
semi-analytic model to predict SFHs for haloes in \dmsim{}.
This is how particle tagging is usually applied in practice. 

\BB{In this stage in the comparison, as we stressed in the introduction,} it is
important to separate differences that arise from particle tagging from those
that arise simply because \galform{} predicts a different SFH to \sphsim{}.
Clearly, having reasonably well-matched models of star formation is a
prerequisite for comparing the amplitudes of density profiles and the balance
between \ins{} and accreted stars. The parameters of \galform{} can be
constrained by comparing predictions derived from cosmological volume
simulations to large observational data sets covering a very wide range of
galaxy scales. Arguably the most important constraints are $z=0$ luminosity
functions \citep[\eg][]{Cole:2000aa}. The subgrid star formation model of
\sphsim{}, however, was not calibrated in this way.  Since differences between
the subgrid models used in \sphsim{} and \galform{} are not relevant to our
tests of particle tagging, the best \galform{} parameters to use in our
comparison are those that most closely reproduce the predictions of \sphsim{},
not necessarily those which satisfy the usual observational constraints. In
practice, we find that the \galform{} model used by C10 (essentially that of
\citeauthor{Bower:2006aa} with refinements to the modelling of dwarf galaxies)
reproduces the stellar mass of the MW analogue in \sphsim{} reasonably well,
but underpredicts the mass of accreted halo stars. This indicates star
formation in low-mass haloes is more strongly suppressed by feedback in the C10
model than in \sphsim{}. We therefore introduce a simple variation on C10 in
which we reduce the value of one parameter, $V_{\mathrm{hot}}$, from $450$ to
$250\,\mathrm{km\, s^{-1}}$ (see \citealt{Cole:2000aa} for the definition of
$V_{\mathrm{hot}}$).  This model, which we refer to as the `weak feedback'
variant, reproduces the mass of the accreted stellar halo in \sphsim{} much
more closely\footnote{Massive haloes, of which we only have one in \sphsim{} by
construction, have very different star formation efficiencies to the haloes
that host dwarf galaxies. Simply varying the global strength of feedback as we
have done here does not take into account this scale dependence and is hence a
crude way of `matching' our SPH and semi-analytic models, but is sufficient for
our purposes.}. In contrast with this variant, we refer to the default
parameters used by C10 as corresponding to relatively `strong feedback'. The
correspondence between individual satellite SFHs in these \galform{} variants
and those in \sphsim{} is close on average but of course not exact. Further
details are given in Appendix~\ref{appendix:galform_sfh}.

Fig.~\ref{fig:c4_main_galform} shows the surface brightness profiles arising
from tagging \dmsim{}, based on SFHs predicted by the two
variants of the C10 \galform{} model, and compares these with the original SPH
star particle profile.  Given the many approximations involved, the weak
feedback model reproduces the SPH results remarkably well, particularly in the
case of the accreted component. The density of the stellar halo in the weak
feedback variant is slightly higher than that of \sphsim{}, which alleviates
the discrepancy in the region $10<r<30 \, \mathrm{kpc}$ identified for the
`transplant-tagging' comparison in the previous subsection. The stellar halo is
also less diverse in this region (lower $N_{\mathrm{sig}}$) in the
semi-analytic realizations, which implies that the contributions of individual
progenitors has changed. The SFHs of satellites are at
least as important as the dynamics of their host subhaloes in explaining the
density profile of the stellar halo at this level of detail.

Finally, it is not surprising that we find more difference in the density of
stars formed \ins{} in the MW analogue than in the accreted component.  As
noted by C10, the physical assumptions used to justify particle tagging are not
expected to hold for this component. It is therefore interesting that the
profile the \ins{} component is reproduced as well as it is\footnote{The
results of C13 demonstrate that this is not a coincidence for the star
formation and assembly history of this particular galaxy. With \stings{}, the
half-mass radius of the \ins{} component is explicitly related to that of the
host DM halo by construction, at least for plausible $\Lambda$CDM SFHs.}, with
similar extent and half-mass radius. The differences between particle tagging
and SPH results for the \ins{} component are well within the range of variation
between predictions for this specific set of initial conditions using different
hydrodynamical schemes \citep{Scannapieco:2012aa}.  \BB{Overall, the
semi-analytic profile for the \ins{} component in \dmsim{} shows a similar
discrepancy with the SPH star particles to that seen in
Fig.~\ref{fig:c4_main_sph_tags}, where we tagged DM particles in \sphsim{}
simulation itself. This suggests that most of the discrepancy is due to an
intrinsic limitation of the tagging scheme, rather than differences in the
gravitational potential or the SFH of the main galaxy between \sphsim{} and
\dmsim{}. If that were true, a different tagging procedure might improve the
agreement even further. In the following section we will explore this idea in
order to better understand why particle tagging performs so well in this
comparison.}

\section{Limitations of tagging fixed fractions of DM by energy rank}
\label{sec:details}

\subsection{Binding energy distributions}
\label{sec:nearest}

\begin{figure}
\includegraphics[width=84mm, trim=0.0cm 0cm 0.0cm 0cm,clip=True]{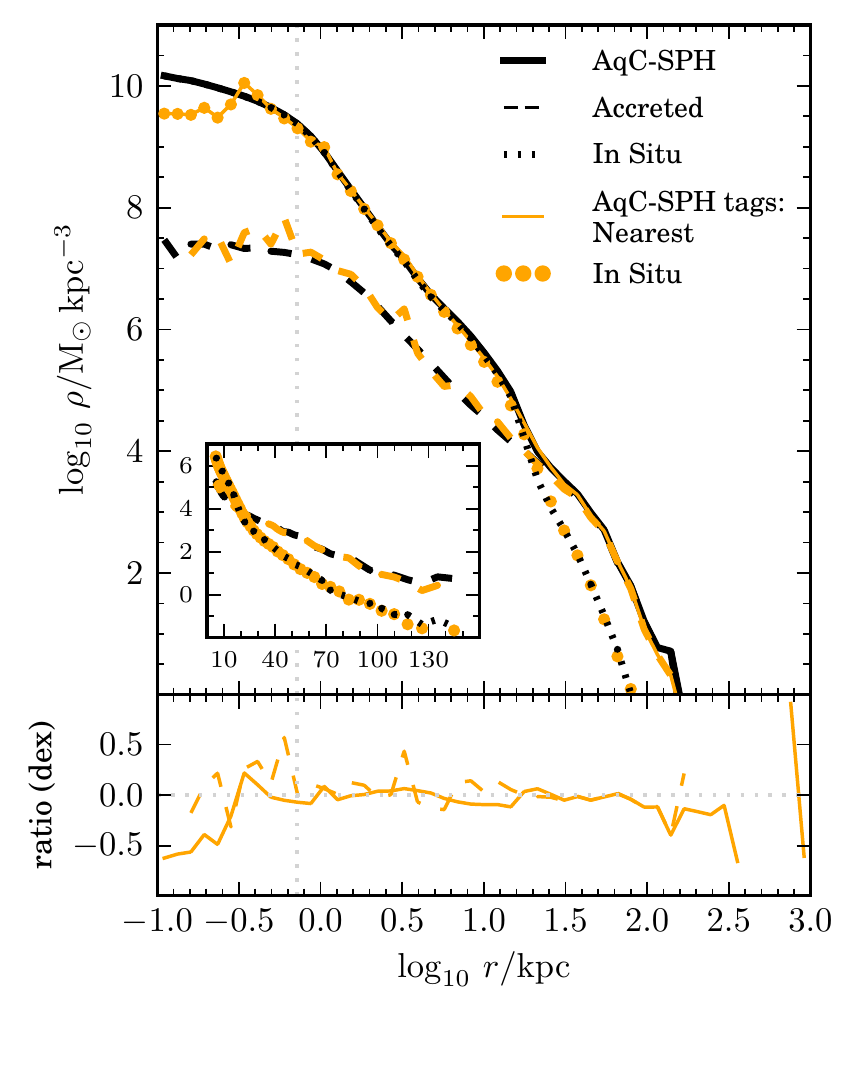}
\vspace{-4em}
\caption{Density profiles for all main halo stars in \sphsim{} (solid),
subdivided into accreted (dashed) and \ins{} (dotted) components. Colours
correspond to star particles (black) and stellar mass carried by tags (orange),
assigned according to the idealized nearest neighbour scheme. The inset shows
the same profiles on a linear scale, with finer binning.  The lower panel
shows the logarithmic ratio of the SPH and tagged star profiles for all stars
(solid) and accreted halo stars only (dashed). A vertical dotted line marks the
force softening scale.}

\label{fig:c4_main_near}
\end{figure}

One of the fundamental assumptions common to all particle tagging
implementations is that it is possible to find, for each newly formed stellar
population, a set of DM particles that have similar phase-space trajectories
over the time-scale of interest (i.e. a Hubble time). Generally speaking,
different implementations assume different forms for the energy distribution of
the stellar population and assign weights to particles in the same DM halo in
order to reproduce those distributions.  For example, the fixed-fraction
\stings{} scheme assumes the stars uniformly sample the \BB{binding energy}
distribution of the most tightly bound region of the potential at the instant
of their formation (i.e. the softened NFW distribution function truncated at a
particular relative binding energy).  In all these schemes, however, the
initial energy distribution of the tags is likely to be a relatively crude
approximation to that of an analogous stellar population in a
hydrodynamical simulation. Moreover, the phase-space trajectories of
star particles are not functions of energy alone. Stars are likely to be formed
on circular orbits, at least in the cold, quiescent discs of MW-like haloes. By
ignoring the angular momentum of DM particles associated with the stellar tags,
only the phase-space excursions of the stars can be approximated, rather than
their actual trajectories.\footnote{\label{footnote:angular_momentum}As C10
note, it would be possible to include a high angular momentum at a given energy
as another criterion in the selection of DM particles. However, that would
greatly limit the number of suitable particles available, because relatively
few DM particles are on circular orbits.}

Using our SPH simulation, we now explicitly test the assumption that a suitable
set of DM particles \textit{can} be found in a scheme based only on
binding energy (putting aside the question of how to find it in practice).  We
do this by searching for sets of DM particles that match the initial binding
energy distributions of each SPH stellar population as closely as possible.
For every individual SPH star particle, we identify a `nearest neighbour' DM
particle by sorting all particles in the same host halo in order of their
binding energy at the snapshot following its formation. We select the first DM
particle with higher rank (lower binding energy) than the star particle as its
`neighbour'. In cases where the star particle is more tightly bound than all
the DM particles in its halo, we select the most bound DM particle. A DM
particle can be selected as the neighbour of more than one star particle, in
which case the associated stellar mass is increased accordingly.  There are no
free parameters in this selection procedure\footnote{Since the search for
neighbours is limited to the time resolution of the simulation snapshots, the
precision of this choice is more limited than it has to be -- for increased
precision the search could be done `on the fly', while the simulation was
running.  Moreover, the phase-space trajectories of tags will not correspond
perfectly to those of their associated star particles even in this idealized
scheme, because the star particle distribution function need not be a function
of energy alone, and because the initial trajectories of star particles may not
be well sampled by DM particles (see point (iv) in the Introduction,
and footnote~\ref{footnote:angular_momentum}). As well as being on
more circular obits}, gas particles may also be more tightly bound than the
most bound DM particle when they are converted to stars, in which case the use
of the most bound DM particle is somewhat arbitrary and its accuracy dependent
on resolution. 

\begin{figure*}

\includegraphics[width=168mm, trim=0.0cm 0cm 0.0cm
0cm,clip=True]{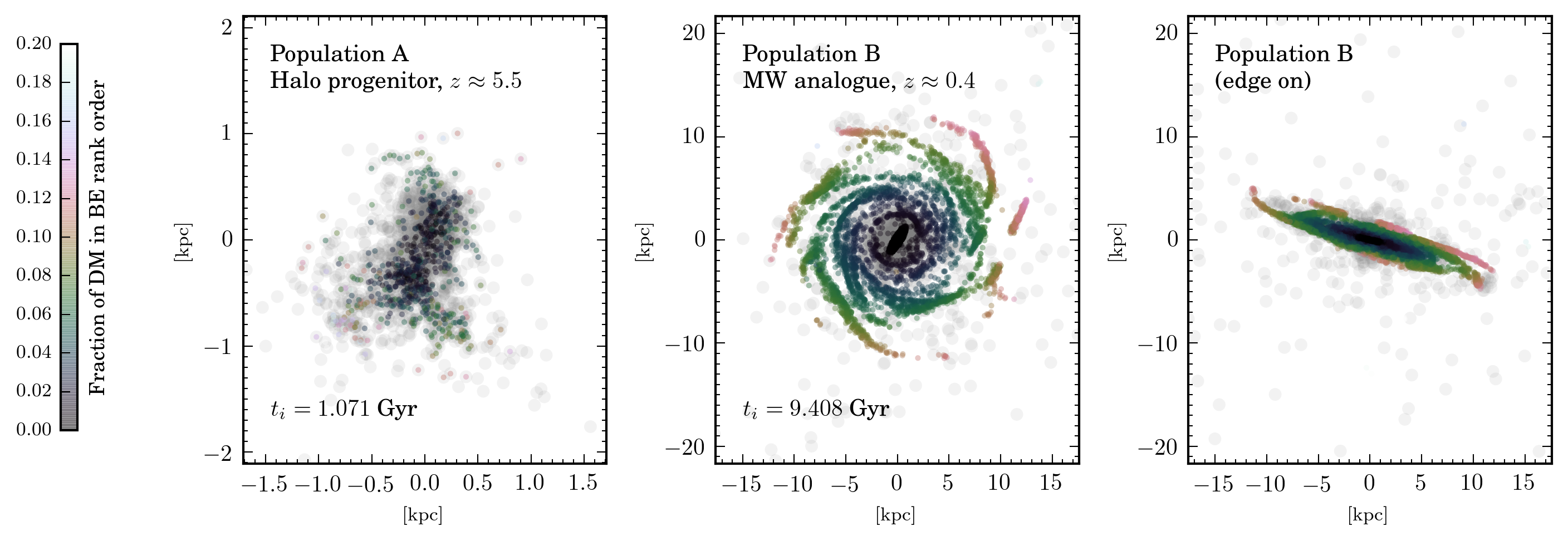} 

\caption{Two examples of the `initial conditions' of single stellar populations
(SSPs) in \sphsim{} at their formation time ($t_{i}$, measured from the big
bang to the present day at $t=13.582\, \mathrm{Gyr}$). \textit{Left:} a
`starburst' population (A) formed at $z\approx5.5$ in one of the larger
progenitors of the accreted stellar halo. The dwarf galaxy host of these stars
is fully disrupted in the main MW analogue halo before $z=0$.  \textit{Centre:}
a different population (B) formed in the disc of the MW analogue galaxy
itself at $z\approx0.4$.  Note the central bulge/bar and associated `ring' of
star formation.  \textit{Right:} an orthogonal projection of population B,
approximately edge-on. Colour indicates the fraction of DM in the host halo
with higher binding energy rank than the particle at $t_{i}$ (see the text).  Grey points are
pre-existing stars in the host at $t_{i}$ (only $1/1000$ of these are shown);
by definition these do not belong to the SSPs A or B and are shown only for
scale.}

\label{fig:c4_diag_near_image} 
\end{figure*}

Fig.~\ref{fig:c4_main_near} shows the stellar halo density profile resulting
from this idealized scheme, analogous to the results in
Figs \ref{fig:c4_main_sph_tags} and \ref{fig:c4_main_galform}. The
correspondence between the tag and star particle representations of the
accreted stellar halo is very close, as in Fig.~\ref{fig:c4_main_galform},
although here the tagged particle distribution is somewhat nosier because the
number of tags is much smaller (roughly one DM tag for each star particle).
More remarkable is the spherically averaged distribution of the tags associated
with \ins{} star formation, which in this case agree equally well with the SPH
result.  The discrepancies at $\lesssim1$~kpc scales are simply due to the fact
that the DM particle softening length is larger than that of the SPH star
particles.  Since the tagged DM particles in this experiment move in a
potential that includes the contribution from the gaseous and stellar discs,
their distribution is mildly oblate. Nevertheless it is surprising that such
good agreement is obtained in the range $1<r<10$~kpc where most \ins{} star
particles are in a thin disc. The agreement here is much better than it is for
the identical potential in Fig.~\ref{fig:c4_main_sph_tags}, \BB{the only
difference being the use of an idealized tagging scheme rather than the
standard fixed-fraction approach}.

We might also expect poor agreement at very large radii, because, in the
\sphsim{} simulation, an \ins{} stellar halo is built up by stars that form in
streams of weakly bound tidally stripped gas \citep{Cooper:2015aa}, rather than
by the outward scattering of stars formed deep in the potential well.  Since
star formation in these stripped gas streams is triggered by local fluctuations
in gas density, it is unlikely to correlate with the total binding energy of
the stripped gas particles. In such cases, the DM particles that are selected
as `nearest neighbours' are much more likely to be smoothly accreted DM than to
be associated with the parent stream of the corresponding gas particle.  In
practice, in our simulation, this effect is not notable -- the tagged \ins{}
stellar profile agrees very well with the SPH simulation even at $\sim30$~kpc.

\subsection{Examples of individual stellar populations}


To understand why the idealized `nearest energy neighbour' tagging scheme
produces a distribution of `tagged' DM particles that agrees so well with the
distribution of SPH star particles in the same simulation, we now examine the
\BB{dynamical evolution} of individual stellar populations, as in \theo{}.

In Fig.~\ref{fig:c4_diag_near_image} we choose two examples of stellar
populations (sets of star particles that form \ins{} in their host halo between
two consecutive snapshots) with very different dynamical histories, labelled
`A' and `B', respectively.  Population A ($M_{\star} = 1.3\times10^7\Msol$)
forms in a low-mass DM halo at $z\sim5.5$ (a lookback time of 12.5 Gyr). This
halo is subsequently accreted by and disrupted within the MW analogue
halo. As a result, population A is phase-mixed into the stellar halo of the
MW analogue by $z=0$.  Population B ($M_{\star} = 1.0\times10^8\Msol$)
forms \ins{} in the MW analogue halo itself at much lower redshift ($z\sim0.4$;
lookback 4.2 Gyr) and can be considered a `MW disc' population. 

The figure shows the projected distribution of star particles in each
population at the simulation snapshot immediately after their formation (this
is the same for all particles in the population, by definition). It is clear
that population B forms within the thin baryonic disc of the MW analogue
(shown approximately face-on and edge-on) whereas population A has a more
amorphous distribution within its initial host. Note the prominent stellar bar
in the inner $\sim2$~kpc of population B, and also the order of magnitude
difference in spatial scale between the two populations. \BB{Clearly, the
definition of a single population in the context of particle tagging does not
correspond to the usual concept of a single star-forming region in the case of
population B, where stars are forming across the entire disc}.  The colour of
each point corresponds to the fraction of DM more bound than a given star
particle (for example, 5 per cent of the DM particles are more bound than the
star particles shown with blue colours). The clear radial gradient in colour
reflects a tight correlation between binding energy and depth in the potential,
even for the centrifugally supported disc of population B.

\begin{figure}
  \begin{center}
\includegraphics[width=80mm, trim=0.0cm 0.2cm 0.0cm 0cm,clip=True]{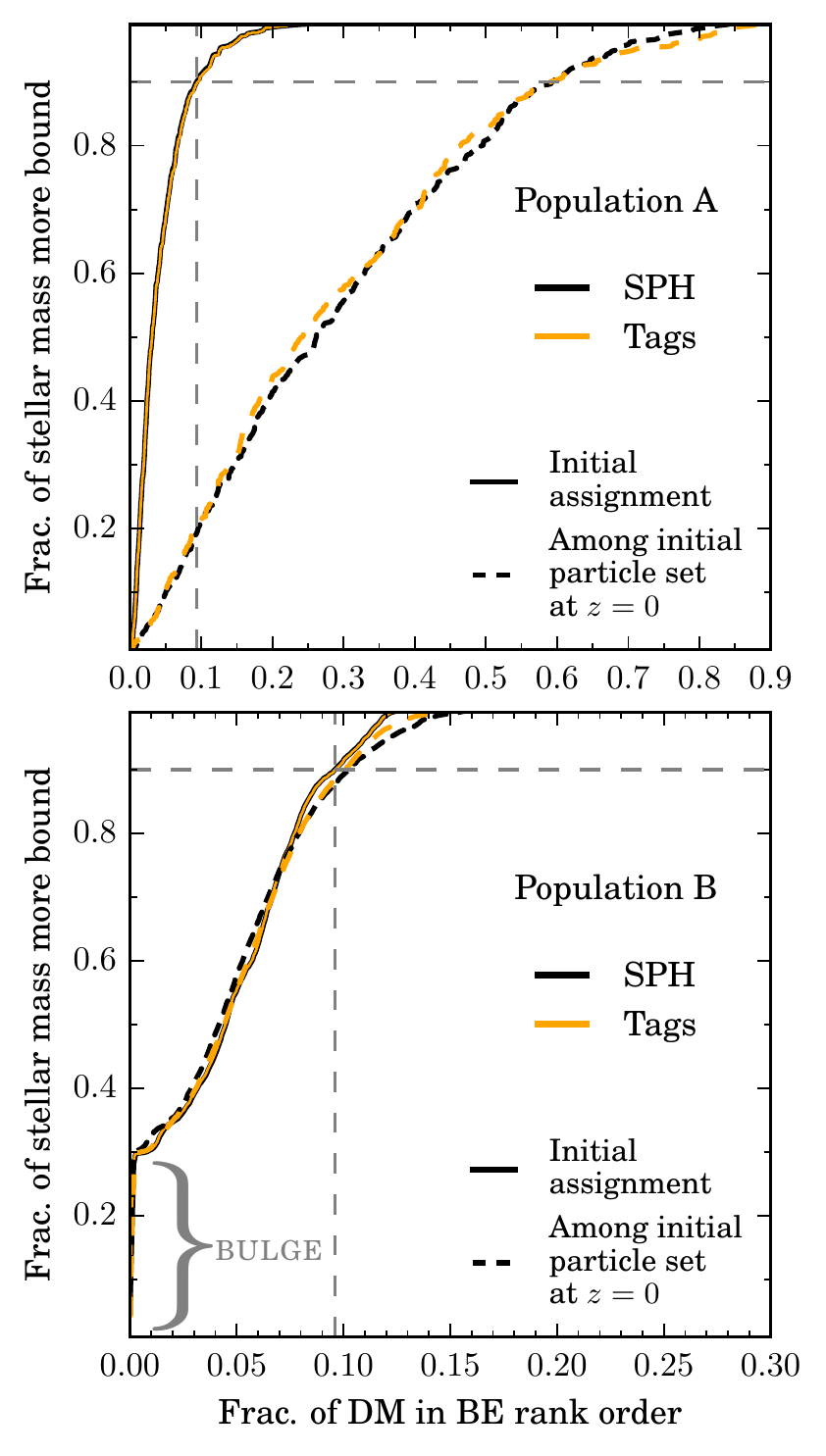}
\end{center}

\caption{Cumulative fraction of stellar mass in the newly formed populations A
and B (Fig.~\ref{fig:c4_diag_near_image}) that is more bound than a given
fraction of the DM in their corresponding host haloes. The horizontal axis
corresponds exactly to the colour scale of the images in
Fig.~\ref{fig:c4_diag_near_image}. The nearest energy neighbour tagging scheme
(orange) applied to DM in \sphsim{} accurately reproduces the distributions of
the actual star particles (black) at the formation time (solid lines;
essentially by construction in this scheme). Vertical dashed lines indicate
$f_{90}$, the fraction of DM enclosing 90 per cent of the stellar mass at
$t_{i}$, an empirical equivalent to $\fmb$ (see the text). The distribution of
the tags still traces that of the star particles at $z=0$ (dashed lines).  We
mark the region corresponding to the nuclear `bar/bulge' of population B in
Figs \ref{fig:c4_diag_near_image} and \ref{fig:c4_diag_near_dens}.}

\label{fig:c4_diag_near_rank}
\end{figure}

In Fig.~\ref{fig:c4_diag_near_rank} we quantify this relationship in more
detail using a distribution directly relevant to particle tagging: the fraction
of newly formed star particles that are more bound than given fraction of the
DM particles in the same halo, when the latter are sorted in rank
order of binding energy. We show this distribution for SPH star particles
(black) and for the DM particles to which we tag their stellar mass in the
nearest energy neighbour scheme (orange). By construction in this scheme, the
distributions of stars and tags are almost identical at the time of tagging.

Of more interest in this idealized case is the correspondence between
star particles and tagged DM particles at the \textit{final} simulation output
time ($z=0$), shown by the dashed lines in Fig.~\ref{fig:c4_diag_near_rank}.
These distributions show the evolution in a `relative' sense (including only
the DM particles that were part of the set available for tagging at $t_{i}$)
rather than an absolute sense (which would include \textit{all} the DM in the
halo at $z=0$).  Relative evolution occurs because tagged DM particles
subsequently diffuse to higher energies, and DM particles that were not tagged
diffuse to lower energies. The $z=0$ distributions for tagged DM particles and
stars are very similar in population A, despite the large time interval and the
complete disruption of the original host halo which `scrambles' the initial
relationship between the binding energy of stars and DM.  The correspondence is
even closer for population B; there is almost no evolution in the `relative'
sense even after 4~Gyr, implying that the DM halo and the galactic disc are
dynamically stable over this period.

\begin{figure}
  \begin{center}
  \includegraphics[width=80mm, trim=0.0cm 0.2cm 0.0cm 0cm,clip=True]{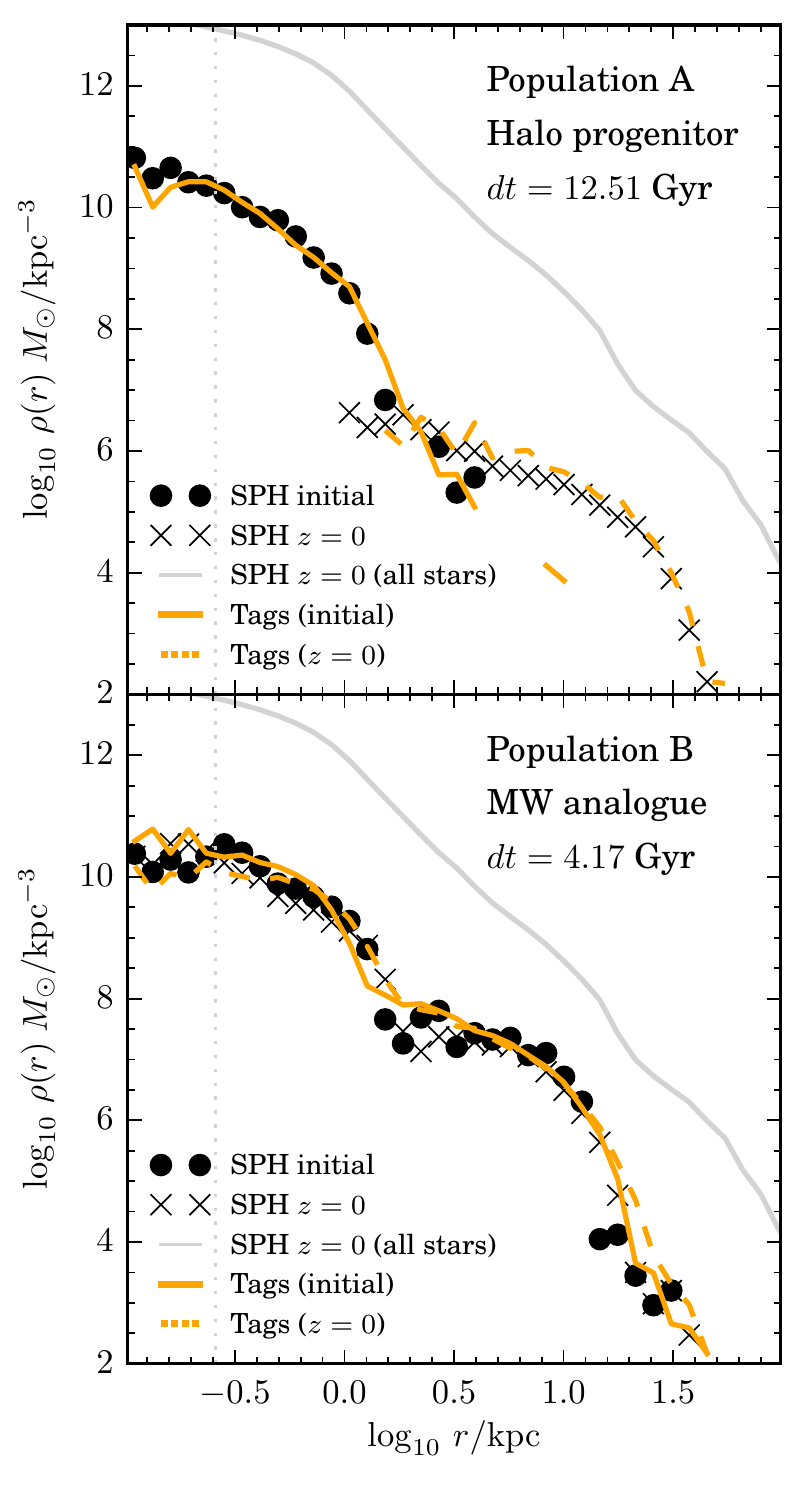}
\end{center}
\vspace{-0.25cm} \caption{Density profiles of star particles (black) and tagged
particles (orange) in the populations shown in
Fig.~\ref{fig:c4_diag_near_image}.  Dots/solid lines correspond to profiles at
the formation time, $t_{i}$. Crosses/dashed lines correspond to the
distribution of the same particles at $z=0$ (the elapsed time, $dt$, is
indicated). In the case of population A (top), the initial profile is centred
on the halo in which it forms, and the final profile is centred on the MW
analogue halo, into which the population is accreted by $z=0$. For scale, the
grey line shows the profile of all stars at $z=0$. Note the transition in
profile B around $\sim1$~kpc, corresponding to the extent of the nuclear `bar/bulge'
region in Figs \ref{fig:c4_diag_near_image} and \ref{fig:c4_diag_near_rank}.}

\label{fig:c4_diag_near_dens}
\end{figure}

Fig.~\ref{fig:c4_diag_near_dens} shows the spherically averaged density profile
of the two populations. Black circles show the initial profile of star
particles, and black crosses their profile at $z=0$.  These can be compared
with the solid and dashed orange lines, respectively, which show the profile of
the corresponding tagged DM particles. For scale, a solid grey line shows the
density profile of \textit{all} star particles in the host halo (the main MW
halo in both these examples) at $z=0$. Although the evolution in the density of
tags and stars does not correspond as closely as their evolution in binding
energy, the differences are still relatively small ($\lesssim 0.5$~dex). As
noted above, population A forms in a dwarf galaxy that is incorporated into the
accreted halo of the `MW' before $z=0$, hence the initial and final profiles
are measured with respect to the centre of the formation halo and the $z=0$ MW
halo, respectively.  For population B, the star particle profile has three
`components' (broadly, the inner bar/bulge, an exponential disc and an \ins{}
halo). These components are reproduced by tagged particles at both the initial
and final times. Remarkably, tagged DM reproduces the spherically averaged
scale length of the SPH disc and bulge and preserve this correspondence over
many Gyr, despite the complexities of the separate star-forming regions and
their very different distribution in configuration space.

We conclude that, given an SPH simulation, it is possible to select sets of DM
particles that trace the evolution of the spherically average density
distribution of star particles in the same simulation to an accuracy better
than a factor of two. This is the case at least for the sub-grid star formation
model implemented by our SPH simulation and holds even for \ins{} stars forming
in a thin disc. When particle tagging is applied in practice, an initial energy
distribution has to be determined a priori, necessarily with some
approximation.  We argue that success or failure in reproducing the
distribution star particles in SPH simulations with tagged DM particles in the
same simulations is almost entirely determined by the accuracy of this
approximation with respect to the true initial energy distribution of star
particles.

\section{What choice of $\fmb$ is appropriate for fixed-fraction tagging?}

In practice particle tagging schemes are applied to simulations that do not
already include a separate dynamical component representing stars, and
therefore have to use simple approximations for the initial energy distribution
of stellar populations. \checkme{For example, the \stings{} scheme assumes}
these distributions can be approximated by those of DM particles selected in
rank order of binding energy from the most bound down to a specified fraction.
The free parameter of the method, $\fmb$, sets the `bias' between the energy
distribution of newly formed stars and the DM of their host halo. This bias is
assumed to be universal, hence `fixed fraction'.  This approximation is
simplistic and it is no surprise that it breaks down in detail for complex star
formation regions dominated by the baryonic potential and having significant
angular momentum, like the MW disc (as illustrated by population A in
Fig.~\ref{fig:c4_diag_near_image}). 

\begin{figure}
\includegraphics[width=84mm, trim=0.0cm 0cm 0.0cm 0cm,clip=True]{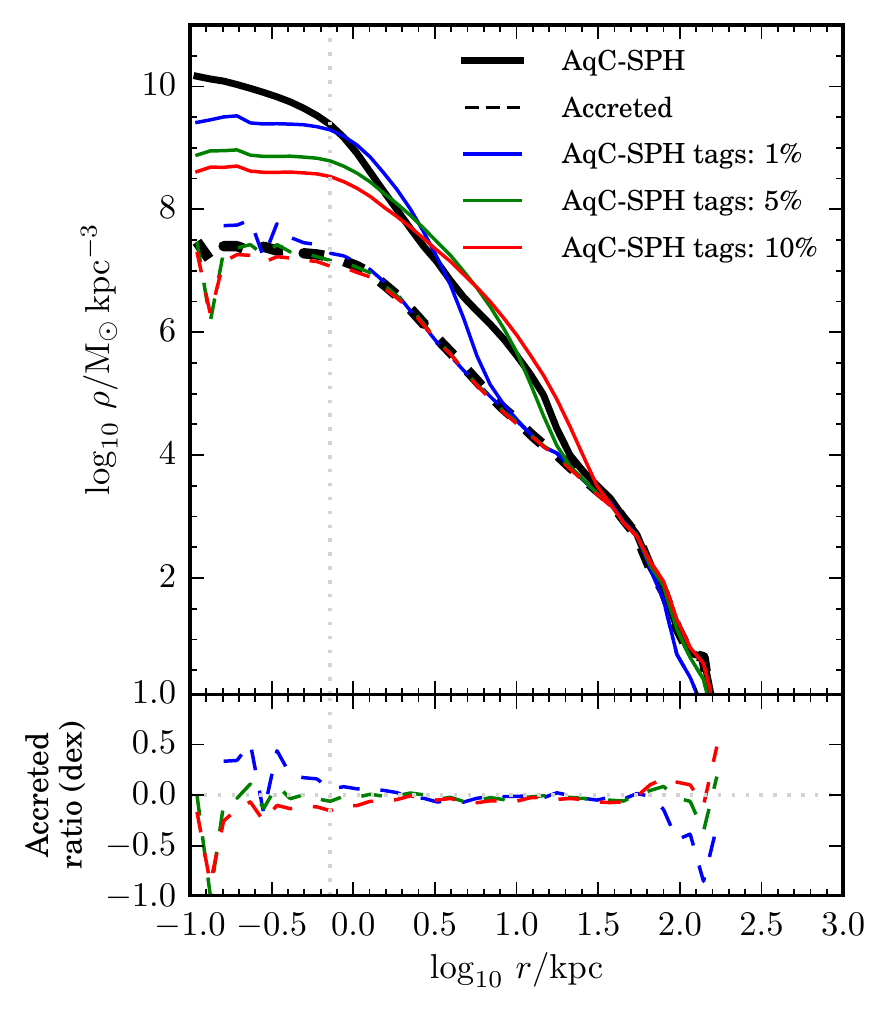}

\caption{Stellar density profile of the main (MW analogue) halo  at
$z=0$ for fixed-fraction tagging based on the SPH SFH with
three different choices of $\fmb$ (see the legend). These are compared to the
result for SPH star particles (black) for all stars (solid) and accreted stars
only (dashed).  Lower panel shows the ratio of accreted stellar mass density in
tagged particles to that in SPH star particles, for each $\fmb$.}

\label{fig:c4_main_std}
\end{figure}

\begin{figure*}
\includegraphics[width=84mm, trim=0.2cm 0.1cm 0.2cm 0.1cm,clip=True]{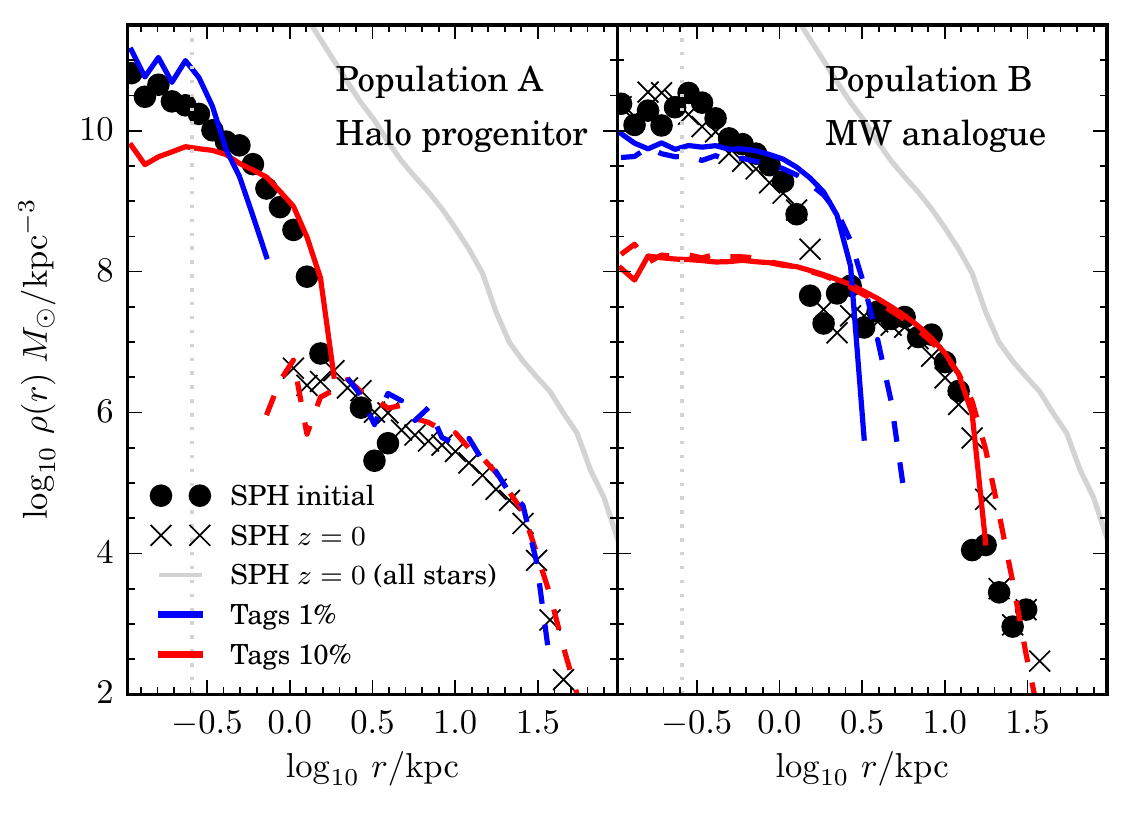}
\includegraphics[width=84mm, trim=0.2cm 0.04cm 0.2cm 0.2cm,clip=True]{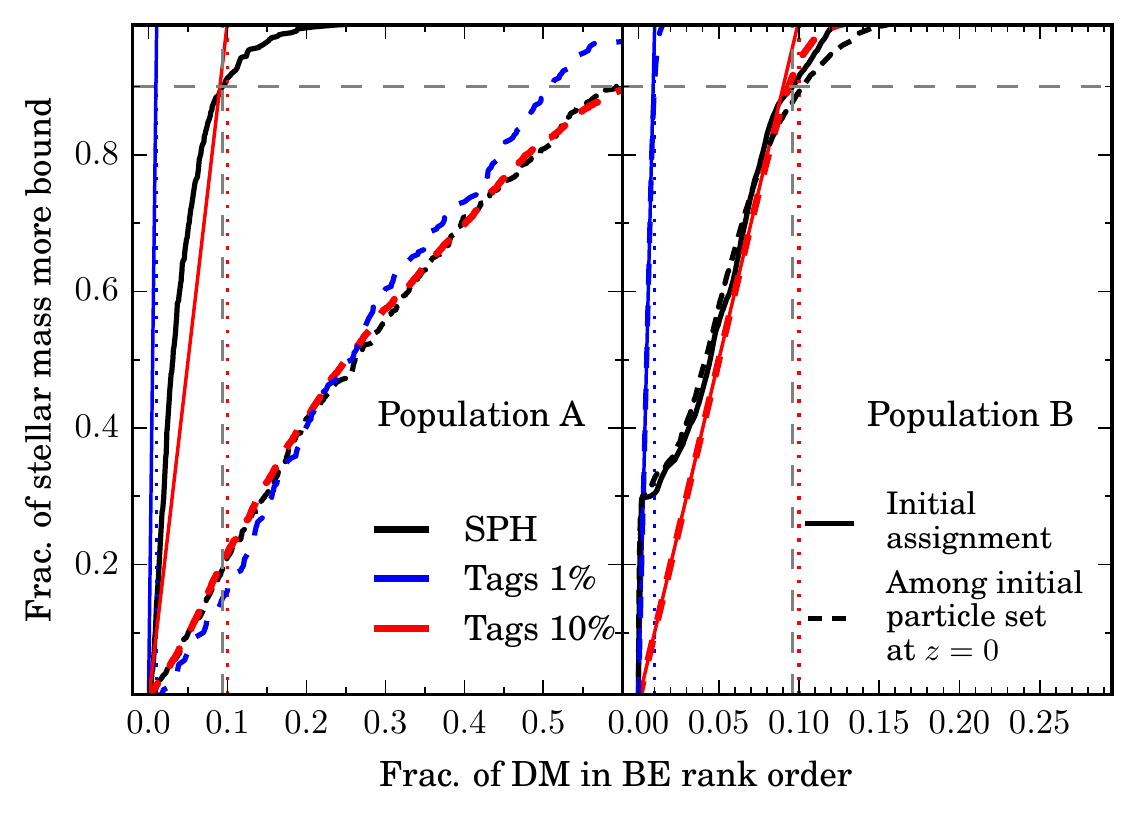}

\caption{Initial- and final-time density profiles (left) and `energy rank
distributions' (right) of single population examples shown in
Figs \ref{fig:c4_diag_near_image} and \ref{fig:c4_diag_near_rank}, here
for fixed-fraction tagging schemes with $\fmb=1$ (blue) and $10$~per~cent (red),
compared with SPH results (black). Dashed vertical lines mark $f_{90}$ and
dotted vertical lines mark $\fmb$ (see the text). Note that fixed-fraction schemes
correspond to linear distribution functions in the right-hand panels (solid
red and blue lines).}

\label{fig:c4_diag_std}
\end{figure*}

Fig.~\ref{fig:c4_main_std} shows how variations of $\fmb$ affect the results of
the SPH-based tagging shown in Fig.~\ref{fig:c4_main_sph_tags}. The good
agreement for the accreted halo distribution is largely insensitive to the
exact choice of $\fmb$. Discrepancies between these four profiles exceed
$\sim10$ per cent only within $r<1$~kpc and beyond $r>100$~kpc. Hence, although
it is sensible to calibrate $\fmb$ with respect to the scale radii of surviving
\ins{} dominated galaxies, results for accreted halo stars are \textit{not}
particularly sensitive to this choice. In most cases, the diffusion of stars in
phase-space associated with the tidal disruption process dominates over small
differences in the structure of the progenitor.  For reasons discussed by C13
and examined in detail by \theo{}, the differences in the `initial conditions'
of their populations are lost by $z=0$. This is not the case for the \ins{}
`disc' populations forming in the very stable central region of the main halo
at low redshift, which consequently show large variations in their shape and
moderate variations in their half-mass radius as $\fmb$ varies from 1 to 10 per
cent. To a lesser extent the same is true for the scale radii of surviving
satellites, which C10 compared to observations to support a value of
$\fmb\sim1$~per cent (further details are given in
Appendix~\ref{appendix:sizes}).

Fig.~\ref{fig:c4_diag_std} illustrates this directly by repeating the analysis
of the individual populations A and B from Fig.~\ref{fig:c4_diag_near_image}
using fixed-fraction tagging with $\fmb=1$ and $10$ per cent. In the case of
population A (the `halo' population), it can be seen that the large initial
differences between the two sets of tags (and between each set and the
corresponding \sphsim{} star particles) are erased by the time the stars and
tags have been mixed into the stellar halo of the MW analogue. In the
case of population B (the `disc' population), neither set of tags evolves
significantly, except for the diffusion of particles in the low binding energy
tail above the initially sharp cut-off energy. The quality of the agreement
between tags and SPH star particles at $z=0$ is therefore dominated by the
initial conditions imposed at the time of tagging. The initial energy
distribution is multimodal, as shown in the previous section, and this clearly
cannot be captured by a single value of $\fmb$. The $\fmb$ scheme corresponds
to a linear form for the `energy rank distribution', whereas this distribution
for actual star particles is at best only approximately linear for the most
tightly bound stars in population A. In population B, comparing the blue and
red lines in Fig.~\ref{fig:c4_diag_std} shows that $\fmb=10$ per cent describes the
bulk of the exponential disc reasonably well, while $\fmb \lesssim 1$ per cent more
closely reproduces the distribution for star particles formed in the compact
nuclear region of the galaxy. 

The most appropriate value of $\fmb$ will clearly differ from galaxy to galaxy,
and from snapshot to snapshot. Putting aside the issue of multiple stellar
populations forming simultaneously in a galaxy, a good empirical approximation
to the optimal value of $\fmb$ for each `aggregate' coeval population in our
SPH simulation can be defined as $\fmb\approx\fnz$, where $\fnz$ is the
fraction of DM in rank order of binding energy enclosing 90 per cent of the
newly formed stars (this definition is illustrated for our two example
populations by the vertical lines in Figs \ref{fig:c4_diag_near_rank}
and~\ref{fig:c4_diag_std}). 

\begin{figure} \includegraphics[width=84mm, trim=0.0cm 0cm 0.0cm 0cm,clip=True]{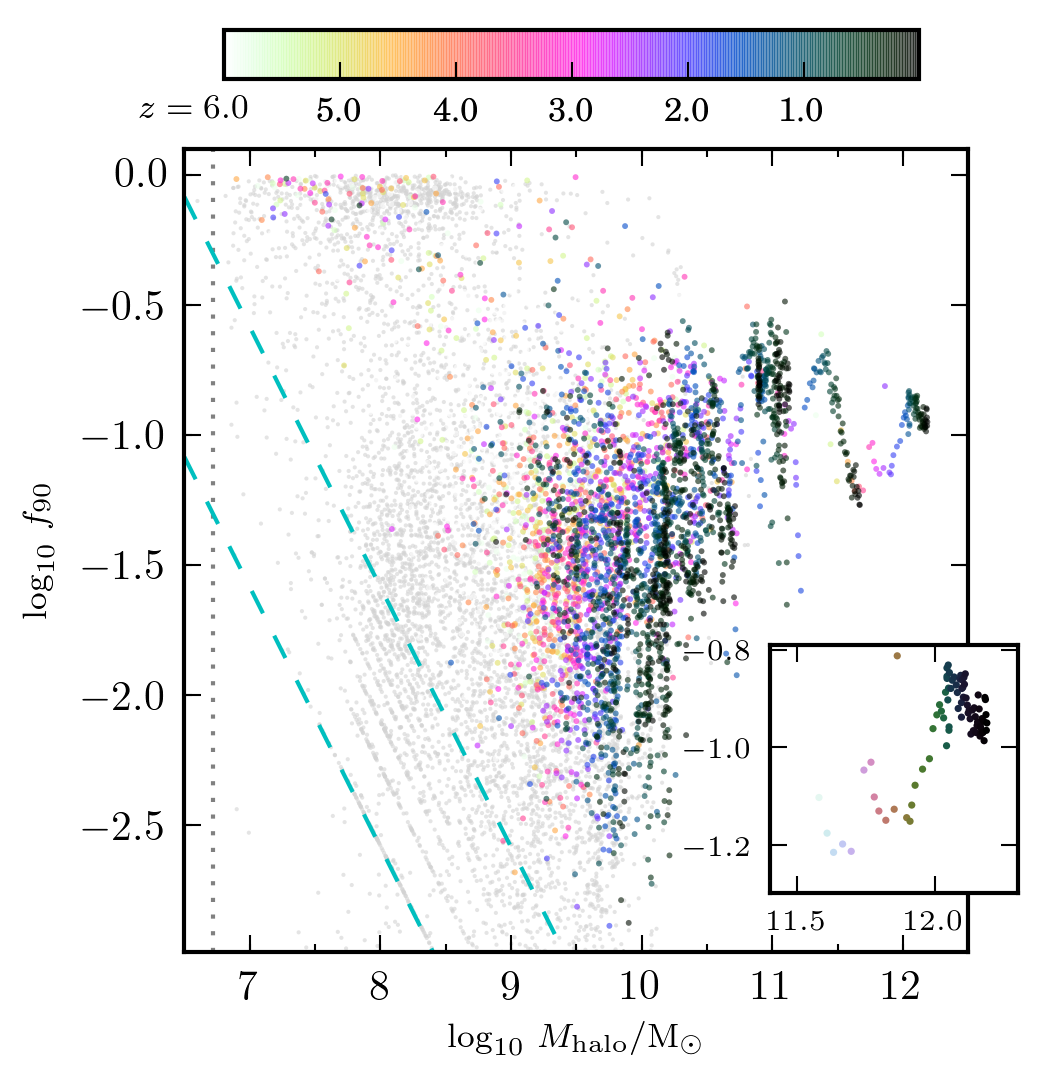}

  \caption{Scatter plot of $\fnz$ (see the text) for SSPs in
  \sphsim{} as a function of instantaneous halo mass measured by
  \subfind{} (main haloes only). Colour coding indicates formation
  redshift: stellar mass is generally a monotonically increasing function of
  time for main haloes. Only star formation events forming five or more star
  particles in haloes with more than 10 DM particles and baryon
  fractions of $\lesssim 20$ per cent (hence excluding spurious clumps of
  baryons) are coloured. Remaining star formation events are shown in grey and
  mostly correspond to small baryon-dominated clumps identified as independent
  haloes.  Dashed blue lines indicate the loci for which $\fnz$ corresponds to
  1 particle (lower line) and 10 particles (upper line). Since the DM particle
  mass is fixed and baryon particle masses vary within a narrow range, the
  smallest nonzero values of $\fnz$ are discretized along lines parallel to
  these loci (there are multiple lines corresponding to small haloes with
  different mixes of the three particle species in \sphsim{}). Inset shows
  evolution for the MW analogue halo only.}

\label{fig:f90_dist}
\end{figure}

Fig.~\ref{fig:f90_dist} plots $\fnz$ for all populations in \sphsim{} against
their stellar mass. A clear sequence of points corresponding to the stable disc
is apparent at $\fnz\sim0.1$, highlighted in the inset panel. For low-mass
populations, there is huge scatter, reflecting the complex nature of star
formation in low-mass DM haloes \BB{(likely in addition to numerical noise)}.
A good understanding of the shape of this distribution under different
star-forming conditions would greatly improve the correspondence between
particle tagging and SPH simulations, although it is not clear that the
complexity of a variable-faction tagging scheme would justified.  Given the
other approximations inherent in the method, our results suggest that the
simple fixed $\fmb$ scheme is probably adequate in most cases where particle
tagging is significantly more efficient than SPH simulations, namely very
high-resolution models of dwarf satellite accretion and computing the
statistical properties of large numbers of galaxies in lower resolution
cosmological simulations. For those applications, only the approximate scale of
the \ins{} component is important, provided energy diffusion is taken into
account as discussed in \theo{} either by `live' tagging (as in \stings{}) or
by explicitly imposing an appropriate distribution function (as in
\citealt{Bullock05} and \citealt{Libeskind11}). For studies of MW-like stellar
haloes, calibration $\fmb$ with reference to the mass--size relation of
\ins{}-dominated galaxies, as in  \cite{Bullock05} and C10, is adequate. In
Appendix~\ref{appendix:sizes}, we discuss this calibration further in light of
the results above. 

\section{Discussion}
\label{sec:discussion}

We can now finally return to the question posed by
Figs \ref{fig:c4_main_sph_tags} and \ref{fig:c4_main_galform}. Given two
simulations from the same initial conditions, one hydrodynamical and the other
using \stings{} (semi-analytic galaxy formation in combination with
fixed-fraction particle tagging), are the theoretical limitations of the
particle tagging approach responsible for the differences we see in the
spherically averaged density profile?  In the case of the simulation we
analyse, this does not seem to be the case.  The discrepancies between SPH and
\stings{} predictions in Fig.~\ref{fig:c4_main_galform} are less than an order
of magnitude and can be explained as the result of differences in the modelling
of star formation, the simplistic form of energy distribution assumed by the
\stings{} fixed-fraction tagging scheme and the use of a universal fraction
parameter in this scheme. None of these are fundamental to the particle tagging
approach and can easily be improved upon. The only clear discrepancy that can
be seen as a clear limitation of particle tagging is the more rapid rate of
disruption of massive subhaloes in our SPH simulation. We conclude that, of the
possible sources of discrepancy listed in the introduction, the most important
limitation of particle tagging is its failure to reproduce this systematic
effect on the orbital evolution of subhaloes in collisionless simulations. We
have not identified the origin of this effect. It is possible (but not yet
proven) that it is due to a modification of the innermost regions of the
gravitational potential induced by the baryons associated with the central
galaxy. Changes in the internal structure of satellites may also be important,
although for the most part these are not evident at the time of infall.
Disc-shocking effects, although important for the overall subhalo
population, seem unlikely to affect the most significant contributors of the
stellar halo.  Remarkably, in the context of the stellar halo, the approximate
and incomplete sampling of the phase-space of hydrodynamic star particles by
collisionless tags appears to be less important than any of these factors.

Our findings give a new perspective on previous comparisons between SPH and
particle tagging models, which we briefly revisit in this section.

\subsection{Energy diffusion}
\label{sec:diffusion}

As discussed in detail by \citet{Le-Bret:2015aa}, the apparently minor
simplification of tagging DM particles at the time at which each satellite
progenitor falls into main halo, as adopted, for example, by \citet{DeLucia08}
and \citet{Bailin:2014aa}, can change the results of comparisons like that
shown in Fig.~\ref{fig:c4_main_sph_tags} and make the agreement between star
particles and tagged particles substantially worse in some cases.  This is
because the tag-at-infall approach does not allow for the prior diffusion of
tagged particles in energy space after the associated stars form, whereas SPH
star particles naturally undergo such diffusion. For that reason, tag-at-infall
schemes are not always good proxies for work using live tagging schemes. Live
schemes such as \stings{} take diffusion into account naturally, although it is
also possible to introduce parameters into the tagging scheme to make a
posteriori corrections for its effects when tagging at infall \citep{Bullock05,
Libeskind11}.

The difference between the tag-at-infall simplification and a live scheme also
depends on the SFHs of satellites in the simulation, for
three reasons.  First, if satellites form the bulk of their stars immediately
before infall then the approximation is obviously more reasonable than if they
form long before.  Second, the basic assumptions required for tagging no longer
hold if star formation in satellites is efficient enough to significantly alter
the density profile in ways that are not captured by a collisionless
simulation.  Third, satellites themselves are the product of hierarchical
mergers and may acquire their own diffuse haloes before infall, but any
distinction between \ins{} and accreted stars \textit{within satellites} is not
captured by single-epoch tagging.  Another, more minor issue is that the amount
of energy diffusion is much reduced if stars are less deeply embedded in their
host potential at the time of star formation (for example, in the \stings{}
scheme, if $\fmb \sim 10\%$ rather than $1\%$). Low numerical resolution has
essentially the same effect \citep[as seen perhaps in][]{DeLucia08} because the
central regions of halo potentials, where diffusion effects are strongest, are
not well resolved to begin with.  In a high-resolution simulation in which star
formation in satellites is inefficient and peaks several Gyr before
accretion on to the proto-MW, this diffusion effect is critically
important. These are the conditions for halo star formation favoured by recent
cosmological simulations. 

\subsection{Comparison with previous tests of particle tagging}
\label{sec:previoustests}

\subsubsection{\citet{Libeskind11}}

\citet{Libeskind11} examine some of the issues above using an SPH simulation of
a Local Group analogue with comparable resolution to ours, alongside a matched
collisionless simulation. They claim that fixed-fraction tagging of satellites
at the time of infall does not reproduce the density profile of the accreted
stellar haloes in their SPH simulation adequately. They advocate an alternative
time-of-infall method, in which the `absolute' potential, $\phi$, of particles
to be tagged must satisfy $\phi \ge \kappa\,\phi_{\mathrm{subhalo}}$ where
$\phi_{\mathrm{subhalo}} = -G M_{\mathrm{virial}} / r_{\mathrm{virial}}$ is
defined at the `edge' of the subhalo immediately before infall. Their optimal
value is $\kappa\sim16$, chosen to best match the density profiles of tagged
particles and stars in their SPH simulation. 

Since the method advocated by \citet{Libeskind11} is applied at the time of
infall, the freedom in choosing $\kappa$ implicitly compensates for diffusion
in energy between the time of star formation and the time of infall, as
discussed by \theo{}.  If the baryonic physics in their simulation
significantly alters the concentration of their potentials, or causes them to
depart from the NFW form, this may explain why they find that fixed-fraction
tagging at infall performs poorly.  If not, their claims in this regard are
hard to understand, because although the \citet{Libeskind11} method requires
the explicit calculation of potential energies, in practice it is essentially
the same as our \stings{} fixed-fraction scheme\footnote{We do not agree with the
statement in section 4 of \citet{Libeskind11} that tagging a fixed fraction of
DM particles by binding energy rank (which they call `relative') is distinct
from (and hence less accurate than) the `absolute' approach they propose.  For
a self-bound, virialized collection of equal mass particles, selecting a fixed
fraction of mass in order of binding energy rank is equivalent to selecting
particles more bound than a fixed multiple of $-G M_{\mathrm{virial}} /
r_{\mathrm{virial}}$, because $\phi$ is a monotonic function of $M(<r)$. At
least part of the discrepancy they discuss is likely to be due to the fact that
their method implicitly corrects for the shortcomings of applying a fixed
fraction scheme at the time of infall, as discussed by \theo{}.}. They report
that this criterion selects about $\sim1$--$3$~per cent of the DM particles
accreted from subhaloes and bound to their three most massive host haloes at
$z=0$. Their criterion therefore appears roughly equivalent to $\fmb\sim1$~per
cent. This is not easy to interpret, however, because, for a simple NFW
profile, the minimum of the potential has
$\phi_{\mathrm{cen}}/\phi_{\mathrm{subhalo}}\le16$ for concentrations
$c_{\mathrm{NFW}}\lesssim46$ \citep[\eg][]{Cole:1996aa}, which implies that,
for a typical mass--concentration relation, no mass should be as tightly bound
as they require in the majority of haloes. Even for a (rather extreme) halo
with $c_{\mathrm{NFW}}=50$, $\kappa=16$ corresponds to only $\fmb\sim0.2$~per
cent.  Libeskind et al. do not recommend a way to apply their technique in
cases where no particles in a subhalo are more bound than their threshold.
Overall, however, the \citet{Libeskind11} study seems to be in broad agreement
with our conclusions and those of \theo{}.

\subsubsection{\citet{Bailin:2014aa}}

\begin{figure} 
  \includegraphics[width=84mm, trim=0.0cm 0cm 0.0cm 0cm,clip=True]{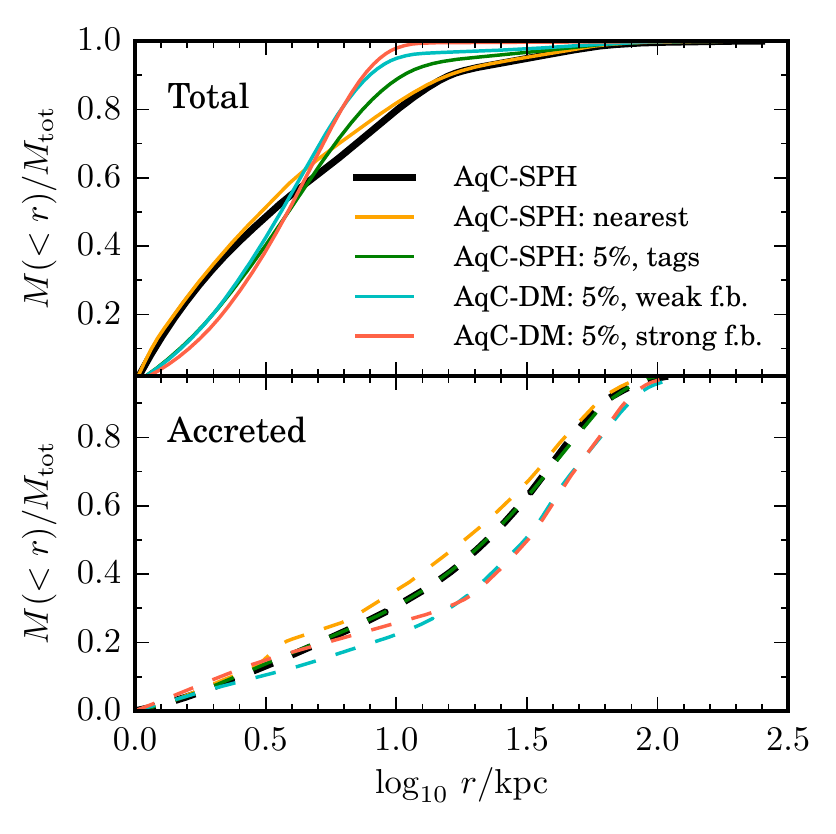}

  \caption{Cumulative mass fraction between 1 and 250~kpc for star particles in
  AqC-SPH (black), and particle tagging results described above for AqC-SPH
  (green and orange) and AqC-DM (blue and red). Upper panel: all stars, lower panel:
  accreted stars only. This figure can be contrasted with fig.~4 of
  \citet{Bailin:2014aa}.}

\label{fig:as_bailin}
\end{figure}

\citet{Bailin:2014aa} present a critique of particle tagging, also in the
context of the MW stellar halo and based on a comparison between
hydrodynamical and collisionless simulations from the same initial conditions.
The discussion and results they present underscore several well-known potential
pitfalls of particle tagging methods, which were noted (and avoided) in the
implementations of \citet{Cooper:2010aa} and \citet{Libeskind11}.  Their work
emphasizes discrepancies in the 3D shape and smoothness of the
stellar halo (the latter quantified by the variance of fluctuations in the
density of halo stars in \CC{broad `zones'} defined in spherical
coordinates).  These measures are \CC{relevant to the interpretation of
observational data \citep[e.g.][]{Helmi:2011aa,Cooper:2011aa,Monachesi:2016ab}}
and could be \CC{sensitive} to differences in how satellites are disrupted
\CC{in hydrodynamical and $N$-body models}.  However, it is not easy to
distinguish the effects of particle tagging on these statistics from other
sources of divergence between SPH and collisionless simulations. We therefore
believe the most direct point of comparison between our work and
\citet{Bailin:2014aa} is their claim that particle tagging artificially reduces
the predicted concentration of accreted stellar haloes because it does not
account for the baryonic contribution to the potential. 

The lower panel of Fig.~\ref{fig:as_bailin} shows the cumulative stellar mass
profiles of accreted stars only, in the region between 1 and 250~kpc, for
several of our model variants. These curves can be compared directly to those
in the lower panel of fig.~4 in \citet{Bailin:2014aa}.  We also include the
corresponding cumulative profile of the total stellar mass (top panel); this
was not shown by \citet{Bailin:2014aa} but is useful for reference here.
\CC{We show these enclosed mass fraction curves only for comparison with
\citeauthor{Bailin:2014aa} because they are not straightforward to interpret.
Compared to the density profiles shown in Fig.~\ref{fig:c4_main_galform}, they
are more sensitive to differences near the centre of the potential ($\lesssim
1$~kpc), which may not be significant in the context of the stellar halo
overall. These curves do not provide information about the absolute density of
each variant at a given radius, only about the relative concentration of their
density profiles. For example, the profiles of the two particle tagging
variants with different feedback strengths (red and cyan lines) appear very
similar in Fig.~\ref{fig:as_bailin}, but very different in
Fig.~\ref{fig:c4_main_galform}.}

\citeauthor{Bailin:2014aa} describe a model, \texttt{SPH-EXACT}, in which each
stellar halo progenitor in their SPH simulation is tagged only once, at the
time of its maximum mass. The sum of the mass of SPH star particles (\ins{} or
accreted) accumulated up to that point is distributed evenly among the 1\% most
bound DM particles\footnote{\citeauthor{Bailin:2014aa} do not state why the
total stellar mass of their \texttt{SPH-EXACT} realization is almost twice that
of the original SPH simulation.}. This can be compared with the two tagging
schemes we apply to \sphsim{} ($\fmb=5\%$, Sec.~\ref{sec:sphtag}; and `nearest
energy neighbour', Sec.~\ref{sec:nearest}). In our case, both of these schemes
predict distributions for the accreted component that are nearly identical to
that of the original star particles (the nearest neighbour results diverge
slightly at $\sim3$~kpc because of a small spike in the density at this radius
which can be seen in Fig.~\ref{fig:c4_main_near}).  \citeauthor{Bailin:2014aa},
in contrast, find their \texttt{SPH-EXACT} halo does not resemble their
star-particle reference model (\texttt{SPH-STARS}). The disagreement they find
is $\sim30$~per~cent at 10~kpc. This difference is even greater than that shown in the
top panel of Fig.~\ref{fig:as_bailin}, which, however, includes \textit{all}
the stellar mass, the majority of which formed \ins{}. We conclude that either
the assumptions of particle tagging are violated much more strongly in the
simulation of \citeauthor{Bailin:2014aa} than in our simulation, or else their
\texttt{SPH-EXACT} model diverges from their \texttt{SPH-STARS} model for
reasons other than the limitations of particle tagging alone.

\citeauthor{Bailin:2014aa} also use a simple linear relation between stellar
mass and halo mass obtained from their \texttt{SPH-STARS} model to carry out
tagging in their collisionless simulation (their \texttt{DM-PAINTED} model).
They find the results of that experiment do not resemble their
\texttt{SPH-STARS} or \texttt{SPH-EXACT} models, and do not agree well even
with the results of the same scaling relation applied to the SPH simulation
(their \texttt{SPH-PAINTED} model). In contrast, we find that tagging AqC-DM
using SFHs from \galform{} (which roughly match those
predicted by AqC-SPH) produces results very similar to those from tagging DM
particles in AqC-SPH according to the SPH SFH. We argued
above that the residual discrepancy is largely due to the (relatively minor)
differences in the SFHs used as input. 

It therefore appears that, in our study, the use of DM particles as proxies for
star particles does not in itself create less concentrated haloes.  Simple
fixed-fraction particle tagging schemes like the C10 implementation of
\stings{} result in a less concentrated stellar distribution overall for the
MW analogue galaxy because they do not account for the presence of
separate star-forming regions seen in our SPH simulation, particularly in the
nucleus of the galaxy.  The apparent concentration difference for the accreted
stars in AqC-DM seen in Fig.~\ref{fig:as_bailin} is hardly notable in
Fig.~\ref{fig:c4_main_galform}, where it is dominated by differences on scales
of $\lesssim2$~kpc.

There are many possible explanations for the differences between our findings
and those of \citeauthor{Bailin:2014aa}. We consider the most important to be:
that they used an SPH simulation that greatly overpredicts the efficiency of
star formation and the resultant baryonic impact on the potential of the
central galaxy and its satellites \citep{Stinson:2010aa,Keller:2015aa}; that
they used a tag-at-infall scheme, which does not allow for diffusion in the
energy of tagged particles between formation and infall (this is particularly
important for SPH simulations with strong feedback, as noted by
\citealt{Le-Bret:2015aa}); and that they did not quantify the effects of
differences between the SFHs used as input to each model in
their comparison.  \citeauthor{Bailin:2014aa} conclude that their findings
motivate the development of more elaborate tagging schemes to overcome the
shortcomings they identify.  Although it would be worthwhile to explore
well-constrained extensions of the simple fixed-fraction approach, our results
suggest that even straightforward implementations like \stings{} are adequate
for many applications of particle tagging to the study of stellar haloes.
\BB{Moreover, the results in Appendix \ref{sec:why_do_individuals_differ}
demonstrate that the most important systematic differences between SPH and
particle tagging do not concern how stars are distributed within satellites
before they are disrupted, but rather the dynamics of subhalo disruption, which
more sophisticated tagging schemes would not be able to address.}

\section{Conclusions}

We have used an SPH simulation to test the assumptions inherent in the
semi-analytic particle tagging scheme of C10 (\stings{}) an efficient but
dynamically approximate method for modelling the phase-space evolution of
galactic stellar haloes. In the case of the simulation we consider, these
approximations appear to be reasonable. We were able to recreate the
spherically averaged density profile of star particles representing the
accreted halo in the SPH version of our simulation by applying \textsc{galform}
and \stings{} to collisionless version of the same simulation.  We also found
that the spatial distribution of \ins{} stellar populations can be reproduced
reasonably well by these schemes, under certain more restrictive conditions.
Our findings support the conclusions of \theo{}, who explored the role of
diffusion in energy space in the comparison of particle tagging schemes to
hydrodynamical simulations. We summarize our results as follows:

\begin{enumerate}

  \item Given a set of recently-formed star particles in an SPH simulation it
    is possible to select subsets of the DM particle distribution in
    the \textit{same} simulation that trace the subsequent phase-space
    evolution of those particles almost exactly, using only their relative
    binding energies.  The `best' outcome possible under a scheme like this
    (approximated here by our `nearest energy neighbour' experiment) is a
    near-exact correspondence between the star particle and tagged particle
    realizations of both the accreted and \ins{} stellar density
    distributions.

  \item More approximate fixed-fraction particle tagging schemes, including the
    \stings{} scheme, reproduce SPH star particle results for the accreted
    stellar halo well  provided that the particles used to trace a particular
    stellar population are selected \textit{at the time when that population
    actually forms} (this is especially true for heavily phase-mixed
    populations). Our results on this point reinforce the conclusions of
    \theo{}.

  \item A fixed fraction tagging scheme applied to a DM only
    simulation (most relevant from a practical point of view) can also yield a
    close match to SPH results for the accreted stellar component,
    \textit{insofar as the orbital evolution of subhaloes agrees between the
    two simulations}.  Differences in debris distributions are smallest for
    heavily phase-mixed populations and streams at large distances from the
    galaxy, and largest for coherent streams produced by the interaction of
    heavily stripped satellites with the inner regions of the potential. The
    spatial extent of the \ins{} component (\eg  its half-mass radius) is also
    recovered by tagging, although its detailed 3D distribution may
    depend on additional factors that are not taken into account even
    approximately by fixed-fraction schemes (especially for complex galaxies
    like the MW; see below). 
    
\item Star formation modelling is the most important `nuisance' factor to
  control for in comparisons between particle tagging and hydrodynamic
  simulations. \textit{In such comparisons, the differences that can be
  attributed to the dynamical limitations of particle tagging only become
  significant when much larger discrepancies arising from the use of different
  star formation prescriptions are eliminated}. Controlling for those
  discrepancies requires either a good understanding of how the star formation
  prescriptions in different models correspond to one another, or else a robust
  procedure for `transplanting' SFHs from a simulation with
  baryons to its collisionless equivalent. Here we find that reducing the
  strength of feedback in our semi-analytic model (which is otherwise identical
  to that used in C10) matches the SFHs in our \sphsim{}
  simulation well and substantially improves the correspondence between tagged
  and star particles, simply by changing when, where and in what
  quantity stars are predicted to form.
  
\item In our simulation, the conditions necessary for good correspondence
  between SPH and particle tagging representations are met for most of the
  significant progenitors of the stellar halo. These include: very little
  long-lasting modification of satellite haloes by contraction or expansion due
  to the motion of baryons; the formation of the majority of accreted stars
  before infall into to the MW-like halo; a lack of strong
  interactions between significant progenitor satellites and the
  baryon-dominated regions of the central potential; and the slow growth of
  that central baryonic contribution, also with limited overall contraction or
  expansion. Particle tagging will naturally provide a worse approximation to
  hydrodynamical simulations in which some or all these conditions are not met.

\end{enumerate}

Our conclusions are limited by the fact that we have only examined one SPH
simulation, as did the similar `comparative' studies of \citet{Libeskind11} and
\citet{Bailin:2014aa}.  Moreover, those earlier studies drew strong conclusions
about the general merits of using particle tagging models to interpret
observational data based on the implicit assumption that specific SPH
simulations were themselves suitable for that purpose. That assumption is hard
to justify without a robust statistical comparison between a cosmologically
representative set of real and simulated galaxies. Our \sphsim{} simulation is a modest
improvement in this respect (it is constrained by the MW satellite
luminosity function and does not suffer rampant `overcooling', demonstrated by
a ratio of stellar mass to halo mass in agreement with abundance matching, a
stable stellar disc and low bulge-to-total mass ratio; \citealt{Okamoto:2010aa},
\citealt{Parry12}).  More recent simulations have made further improvements
with regard to large-scale constraints on galaxy formation
\citep[\eg][]{Sawala:2016aa}.

Hydrodynamical models remain computationally expensive and subject to large,
poorly constrained uncertainties in their `subgrid' recipes, which can easily
overwhelm the advantage of dynamical self-consistency. So long as this remains
the case, our results suggest further tests of particle tagging using larger
samples of haloes from a wide variety of hydrodynamical schemes would be
worthwhile, and motivate the investigation of some improvements to the
methodology. Specifically: 

\begin{enumerate}

  \item  Controlled numerical experiments would be helpful to determine how
    important dynamical differences in the disruption of satellites are for
    applications of particle tagging, in isolation from the many uncertainties
    involved in cosmological hydrodynamical simulations and star formation
    physics. For example, it would be useful to quantify the fraction of
    significant halo progenitors that interact strongly with regions of the
    potential dominated by baryons (i.e. with the disc, in the MW case)
    and how exactly the orbits of these systems differ between simulations with
    and without self-consistent hydrodynamics.

\item Further constraints on the modification of real galactic potentials by
  star formation and feedback would help to inform judgements about particle
  tagging.  There is an ongoing debate on this point in the theoretical
  literature and the observational situation is also uncertain \citep[for a
  recent summary see \eg][]{Oman:2016aa}. For a given star formation model
  applied to an MW-like system, it would be useful to quantify how common
  heavily modified satellite galaxies are among typical sets of halo
  progenitors. If semi-analytic models could be used to predict the degree of
  baryonic modification to satellites, a small number of strongly modified
  satellites could at least be flagged and treated with appropriate caution in a
  subsequent particle tagging analysis.

  \item In the context of fixed-fraction tagging schemes, the most appropriate
    fraction of most-bound particles varies from population to population. This
    suggests a refinement to the scheme in which the tagged fraction can vary
    based on other physical parameters such as galaxy size and the mass of
    newly formed stars.  However, this would require more free parameters and
    may be unnecessarily complex for many applications of particle tagging.
    Likewise, any single-fraction scheme cannot reproduce SPH results for in
    situ stars as well in cases where multiple populations with different
    intrinsic binding energy distributions form in the same halo at the same
    time.  Information about the relative star formation rates in different
    regions of the central potential could be obtained from the underlying
    semi-analytic model and used in a more complex tagging scheme to construct
    a more accurate distribution of stellar binding energies.

  \item Collisionless simulations that account in some way for the baryonic
    contribution to the host galaxy potential (for example by adding a
    smoothly growing disc potential) would likely perform even better in
    comparisons against SPH simulations.  This was the spirit of the approach
    in \citet{Bullock05} and could be greatly improved on with modern numerical
    techniques \citep[\eg][]{Lowing:2011aa}. When implementing a scheme like
    this, it will be important to ensure that the density, radial extent,
    stability and growth rate of baryonic components of the potential satisfy
    observational constraints \citep{Aumer:2013aa,Aumer:2013ab}.

  \item In some hydrodynamic simulations (including our \sphsim{}) the \ins{}
    halo is formed mostly from gas stripped from massive satellites -- the
    same satellites whose stars contribute the bulk of the accreted stellar
    halo \citep{Cooper:2015aa}.  In such a scenario, particle tagging might be
    adapted to approximate the formation of \ins{} stars in streams of
    stripped gas, provided those streams are almost `ballistic' (i.e. they form
    soon after the gas is stripped, such that the kinematics of the stars are
    dominated by the orbital motion of their parent satellite rather than
    hydrodynamic interactions). For example the fraction of recently unbound
    gas converted to stars after stripping could be estimated and stars tagged
    to specific recently unbound collisionless particles.

\end{enumerate}
 
With or without these improvements, particle tagging models are intended as an
approximation and consequently have very clear dynamical limitations. The
results we have reported suggest to us that no galaxy formation theory is
sufficiently well constrained at present to make those limitations more
important than differences between subgrid (or semi-analytic) star formation
recipes.

\section*{Acknowledgements}

We thank the anonymous referee for their careful reading of our manuscript. APC
is supported by a COFUND/Durham Junior Research Fellowship under EU grant
267209 and thanks Andrew Benson for discussions that led to this work.  CSF
acknowledges ERC Advanced grant 267291 `COSMIWAY'. APC, SC and CSF acknowledge
support from STFC (ST/L00075X/1). This work used the DiRAC Data Centric system
at Durham University, operated by the Institute for Computational Cosmology on
behalf of the STFC DiRAC HPC Facility (\url{www.dirac.ac.uk}). This equipment
was funded by BIS National E-infrastructure capital grant ST/K00042X/1, STFC
capital grants ST/H008519/1 and ST/K00087X/1, STFC DiRAC Operations grant
ST/K003267/1 and Durham University. DiRAC is part of the National
E-Infrastructure. We acknowledge use of \texttt{matplotlib} \citep{Hunter:2007}
and NASA's Astrophysics Data System Bibliographic Services.

\bibliographystyle{mnras} 
\footnotesize
\bibliography{astro_master} 
\bsp
\normalsize

\appendix

\section{Using transplanted SPH SFHs as the basis for
particle tagging tests}
\label{sec:why_do_individuals_differ}

In Section~\ref{sec:direct_transfer} we stated that transplanting SFHs from an
SPH simulation to its DM only equivalent is necessarily approximate. Some
divergence is almost unavoidable, which may be physically meaningful in some
cases and stochastic in others. Detailed tests of particle tagging that use
this approach require careful analysis of such divergence for significant
progenitors of the accreted halo in order to isolate effects that are directly
attributable to the tagging technique. In Fig.~\ref{fig:c4_main_galform} we
showed that the distribution of accreted star particles in our SPH simulation
can be reproduced reasonably well by applying particle tagging to a
collisionless simulation from the same initial conditions, using transplanted
SFHs for the 10 most massive progenitors. In this appendix we analyse each of
these progenitors individually to give more insight into the correspondence
between our SPH and collisionless simulations. This analysis raises questions
beyond the scope of our paper, so we present it mainly to highlight the
uncertainties involved and to suggest directions for future work.

\subsection{The most massive halo progenitor}
\label{sec:tricky_satellite}

\begin{figure}
  \includegraphics[width=84mm, trim=0.0cm 0cm 0.0cm 0cm,clip=True]{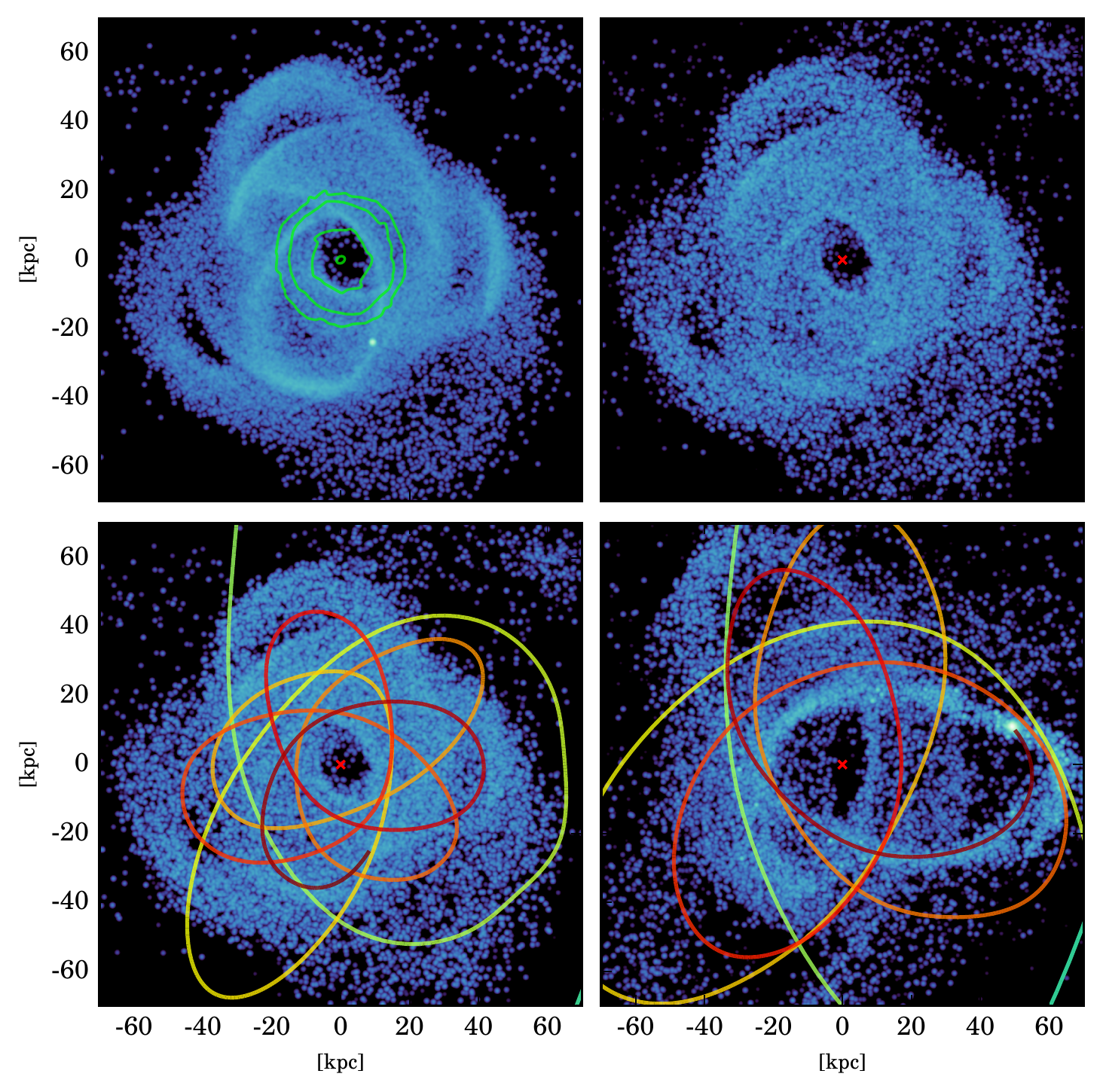}
  
  \caption{Surface density of debris from the `trefoil stream' (the most
  massive halo progenitor) in four realizations. Clockwise from top left:
  \sphsim{}, \sphsim{} with 1 per cent tags, \dmsim{} with 5 per cent tags and
  \sphsim{} with 5 per cent tags. The centre of the main halo is marked by a
  red cross.  The trajectory of the progenitor is shown in the lower panels,
  with increasing time running from green to red.  Contours of the central disc
  surface density are shown in grey in the \sphsim{} panel (top left),
  corresponding to $\log_{10}\Sigma/\mathrm{M_{\sun}\kpc^{-2}} =5$, $6$, $7$
  and $8$.}

  \label{fig:gallery_streams}
\end{figure}

\begin{figure}
  \includegraphics[width=84mm, trim=0.0cm 0cm 0.0cm 0cm,clip=True]{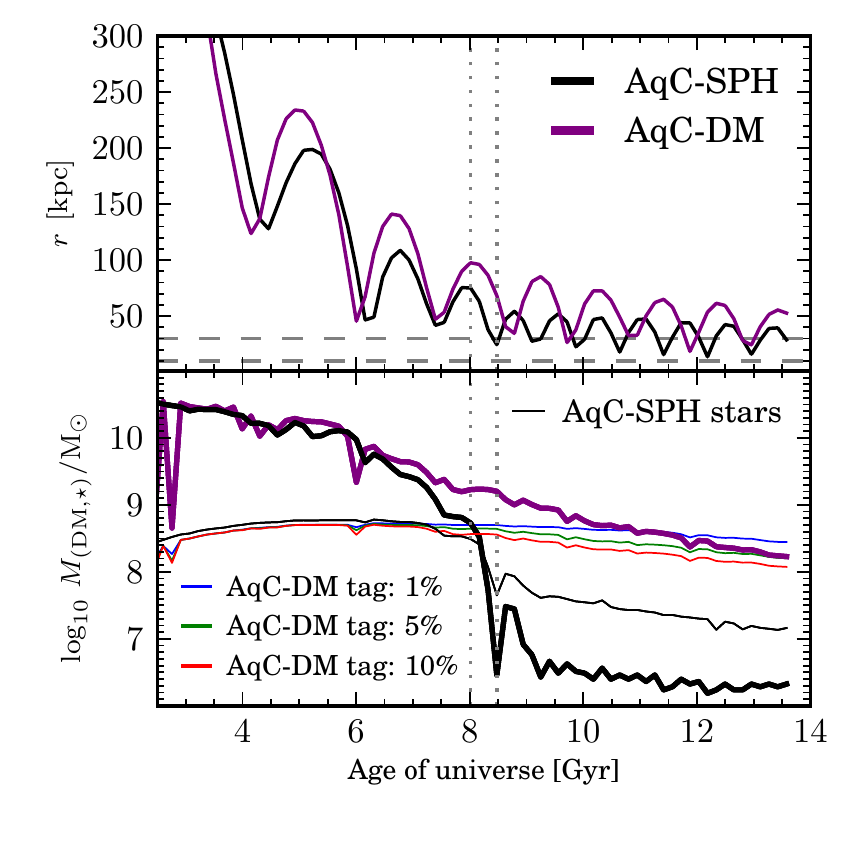}

  \caption{Galactocentric radius (upper panel) and bound mass (lower panel) of
  the trefoil stream progenitor (see the text and Fig.~\ref{fig:gallery_streams})
  with age of the universe, starting from the time of infall into the main
  halo, for \sphsim{} (black) and \dmsim{} (purple). In the lower panel, thick
  lines indicate DM mass and thin lines stellar mass. Thin blue, green and red
  lines in the lower panel correspond to the bound stellar mass predicted by
  tagging of \dmsim{} with $\fmb=1$, $5$ and $10$ per cent respectively.}

  \label{fig:trefoil_orbit}
\end{figure}

We begin with an example that illustrates in detail the case of one stellar
halo progenitor for which the particle tagging results from \dmsim{} do not
correspond well to the star particles in \sphsim{}.  This progenitor is
interesting in several respects, some of which have already been described by
\citet[][section 6]{Parry12}.  At $z=0$, it has been heavily stripped of both
stars and DM, and is responsible for the most striking coherent feature in the
stellar halo. We refer to this feature as the `trefoil stream', on account of
its morphology in the top left-hand panel of Fig.~\ref{fig:gallery_streams} (these
streams lie roughly in the plane of the \sphsim{} disc).  Secondly, this halo
(the `trefoil progenitor') has survived as a satellite of the MW
analogue for $\sim8$~Gyr, undergoing many apocentric passages on a decaying
orbit. The evolution of satellites on orbits like this, and the streams they
produce, should be particularly sensitive to the shape, orientation and
depth of the gravitational potential in \sphsim{}, including the baryonic
contribution absent in \dmsim{}.  Finally, at $z=0$ this object is bound by its
remnant \textit{stellar} mass rather than by DM; we do not expect good
agreement with \dmsim{} in this case where the binding energy of the stars is
critical to the survival of the satellite.  Fig.~\ref{fig:gallery_streams} also
shows the stream as predicted by fixed-fraction tagging of DM particles in
\sphsim{} (top right and bottom left) and tagging of \dmsim{} based on the
transplanted \sphsim{} SFH.  The \dmsim{} version shows
thiner streams, with fewer, wider orbits and a more prominent remnant core at
$(x, y) \sim (50,10)$.

Fig.~\ref{fig:trefoil_orbit} quantifies some of the features that distinguish
the trefoil progenitor from other satellites, comparing its radial position
(relative to the main halo centre) and mass evolution between \sphsim{} and
\dmsim{}. The trajectories diverge around the time of the third apocentre, with
the progenitor in \sphsim{} subsequently having a shorter period and a more
rapid decay than its \dmsim{} counterpart. This divergence seems to be
associated with catastrophic mass loss in \sphsim{} between the third apocentre
and fourth pericentre. The \dmsim{} satellite loses mass more gradually.  The mass
still bound to the satellite at the present day is similar in \sphsim{} and
\dmsim{}, as are the relative orbital phases, despite the fact that the SPH
version passes through two pericentres more than the DM version.  The most
likely reason for the divergence, other than stochasticity, is interaction
between the SPH satellite and the baryons concentrated at the centre of the
potential (the disc and bulge). In the upper panel of
Fig.~\ref{fig:trefoil_orbit}, dashed grey lines are drawn at $10$ and
$30$~kpc, corresponding to disc surface densities of $\sim7$ and $\lesssim4$
$\mathrm{M_{\sun}\kpc^{-2}}$ respectively.

\begin{figure}
  \includegraphics[width=84mm, trim=0.0cm 0cm 0.0cm 0cm,clip=True]{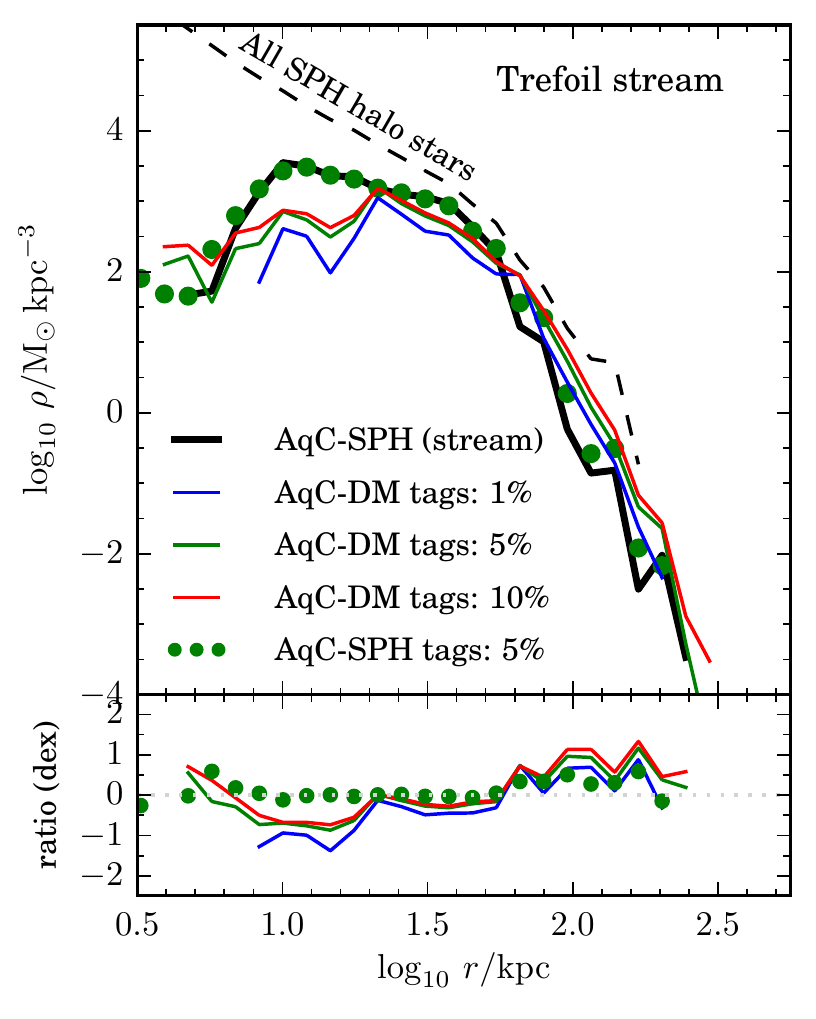}

 \caption{Surface density of stars in the `trefoil stream' (see
 Fig.~\ref{fig:gallery_streams}) at $z=0$ in \sphsim{} (black). Green dots show
 the profile recovered by tagging ($\fmb=5$~per~cent) in \sphsim{}, using the SFH of
 the SPH star particles. Blue, green and red lines show profiles recovered by
 using this same SFH as the basis for tagging of the matched halo in \dmsim{},
 with $\fmb=1$, $5$ and $10$~per~cent respectively. A grey dashed line shows the
 density of all accreted star particles in \sphsim{}. The lower panel shows the
 logarithmic ratio of each curve relative to the SPH result. Note the reduced
 range of radius and shift in density scale relative to previous density
 profile plots for the main halo.}

\label{fig:trefoil_compaison}
\end{figure}

Fig.~\ref{fig:trefoil_compaison} shows the spherically averaged stellar mass
density profile of stars from the trefoil stream progenitor at $z=0$. Tagging
with $\fmb=5$~per~cent in \sphsim{} itself (shown in the lower left panel of
Fig.~\ref{fig:gallery_streams}) results in a close match to the star particle
profile, consistent with the good overall agreement shown in
Fig.~\ref{fig:c4_main_std}. 

When we use the \sphsim{} SFH for this object as the basis
for tagging of the matched satellite in the \dmsim{} simulation, the agreement
is clearly worse (this result is not sensitive to the exact value of $\fmb$).
The predicted density is lower by an order of magnitude at galactocentric radii
below $\sim 20$~kpc, and higher by a similar factor beyond $\sim100$~kpc. This
is readily understood by the differences in orbital evolution shown in
Fig.~\ref{fig:trefoil_orbit}. The mass of stars tagged to the bound core of the
trefoil progenitor in \dmsim{} is shown for $\fmb=1$, $5$ and $10$ per cent.
These indicate an increasing mass-to-light ratio approaching $\sim1$ at the
present day, with the consequence that the stellar mass in the stream (as
opposed to the progenitor) is sensitive to $\fmb$ (increasing by roughly a
factor of 3 as $\fmb$ varies from $1$ to $10$ per cent).

\citet{Parry12} show that the SFH of this satellite is
dominated by an extreme peak associated with rapid gas dissipation following an
early, low mass ratio merger with another halo. This starburst leads to a
cusped stellar density profile and  `explosive' feedback which unbinds a large
fraction of the remaining gas in a short time. The outcome is a
baryon-dominated central cusp and a corresponding low-density DM core.  As
discussed by \theo{}, particle tagging is a poor approximation in cases where
feedback significantly alters the overall density profile. The abnormally low
central density of DM after this event in \sphsim{} is not reproduced in
\dmsim{}, which may then explain why it does not reproduce the rapid loss of DM
relative to stars in the fourth pericentric approach seen in \sphsim{}.  From this
single example, it is hard to divide blame between the abnormal DM density
profile and differences in the interaction with the main halo, since each
reinforces the effects of the other.

In summary, despite the considerable differences in the evolution of the
trefoil progenitor in \sphsim{} and \dmsim{}, we find the distribution of its
debris to be similar overall. The differences we see illustrate the divergence
between results from SPH and DM particle tagging that can be expected in cases
where stellar haloes are dominated by individual `atypical' objects. Satellites
are `atypical' in this case if either feedback significantly alters their
phase-space density, or they interact with a strongly modified central galactic
potential. Further work with SPH simulations is required to understand the
frequency of such cases and their dependence on other aspects of the models. In
our particular simulation, divergence between different realizations of the
most massive halo progenitor (which is atypical on \textit{both} of the above
counts) does not affect the conclusions drawn from the spherically averaged
halo density profile as a whole. The `bias' of particle tagging is not
negligible, but neither is it catastrophic. That bias could, however, alter
other conclusions, for example regarding the extrema of surface brightness
features that might be detected by stream-finding algorithms.

\subsection{Other massive progenitors}

Figs \ref{fig:examples_BCDE} and~\ref{fig:examples_FGHI} show more examples of
massive halo progenitors matched between \sphsim{} and \dmsim{}. These examples
highlight some general systematic differences between the two simulations. As
in the preceding section, the results for \dmsim{} are obtained with
`transplanted' SFHs from \sphsim{} and a 5~per~cent fixed fraction
tagging scheme.  

In some cases the density profiles (leftmost column) of stars accreted into the
main halo by $z=0$ are better matched between \sphsim{} (black solid line) and
\dmsim{} (green solid line)  than in the case of the trefoil progenitor. Only one
case, progenitor F, clearly shows greater discrepancy.  This may be because
progenitor F is disrupted less than $1$~Gyr after infall in \sphsim{} (after
only one pericentre) but survives for $\sim6$~Gyr (six pericentres) in \dmsim{}.
Taking all these examples together, the most obvious systematic difference is
that most progenitors are disrupted somewhat more quickly in \sphsim{} than
they are in \dmsim{} (progenitor C being the only exception).  Only in the case
of progenitor F, however, does this have an obvious effect on the distribution
of the resulting halo stars at $z=0$. Progenitor B is a counterexample, being
similar to C in many respects, having greater divergence in mass-loss rate, and
yet having little discernible discrepancy in the profile of its debris.

It is beyond the scope of this paper to explore why exactly the orbits of
massive satellites diverge in this way between \sphsim{} and \dmsim{}. One
possibility, noted above, is that it is the result of interaction with a
central potential heavily modified (contracted and made more spherical) by
baryonic effects. The mass-loss rates shown in Figs \ref{fig:examples_BCDE}
and~\ref{fig:examples_FGHI} (lower middle panels) suggest another mechanism
related to the to the rapid removal of gas from the progenitors (magenta lines)
most likely by ram pressure as they pass through the densest regions of the
halo \citep{Arraki:2014aa}. It is clear (in particular for B, C and perhaps G)
that gas can be removed more rapidly and essentially disappear after few
orbits.  The total removal of gas seems to correlate with the onset of
divergence in orbit and mass-loss rate of DM between \sphsim{} and \dmsim{}.
Other stochastic effects are possible; for completeness, we note that
satellites often arrive in weakly bound groups of DM haloes with and without
stars \citep[\eg][]{Li:2008aa}.  Small changes in the interactions among
members of these groups before and after infall could be another source of
stochasticity for satellite orbits in cosmological simulations and are
extremely difficult to control for.

The density profiles of debris at $z=0$ (left) and the intact progenitor at the
time of infall (right) also show the predictions of the tag-at-infall approach
used by other studies of particle tagging (dotted green and cyan lines; see
Section~\ref{sec:diffusion}).  Tagging at infall in these examples does not
show much difference compared to the results of live tagging because feedback
in \sphsim{} is relatively weak overall,  as discussed by
\citep{Le-Bret:2015aa} (furthermore, these examples neglect any stars accreted
by the satellite -- including both \ins{} and accreted stars in a single
assignment is another source of inaccuracy in the tag-at-infall approach). The
clear counterexample is the trefoil progenitor, which, as discussed above, has
a small DM core at infall, the innermost part of the galaxy being dominated by
stars. In this case, at infall, the most bound 5~per~cent of the DM particles in
\sphsim{} are good proxy for the distribution of the innermost star particles,
somewhat better even than the live tags in the same simulation, although they
truncate at $\sim1$~kpc whereas star particles (and live tags) extend to
$\sim10$~kpc.  Notably, the equivalent set of the 5~per~cent most bound particles in
\dmsim{} are \textit{not} a good proxy for the \sphsim{} star particles.  Live
tagging of \dmsim{}, on the other hand, results in a reasonable match to
\sphsim{}. This is the only one of our examples to show such complex behaviour.

\begin{figure*}
  \includegraphics[width=150mm, trim=0.0cm 0cm 0.0cm 0cm,clip=True]{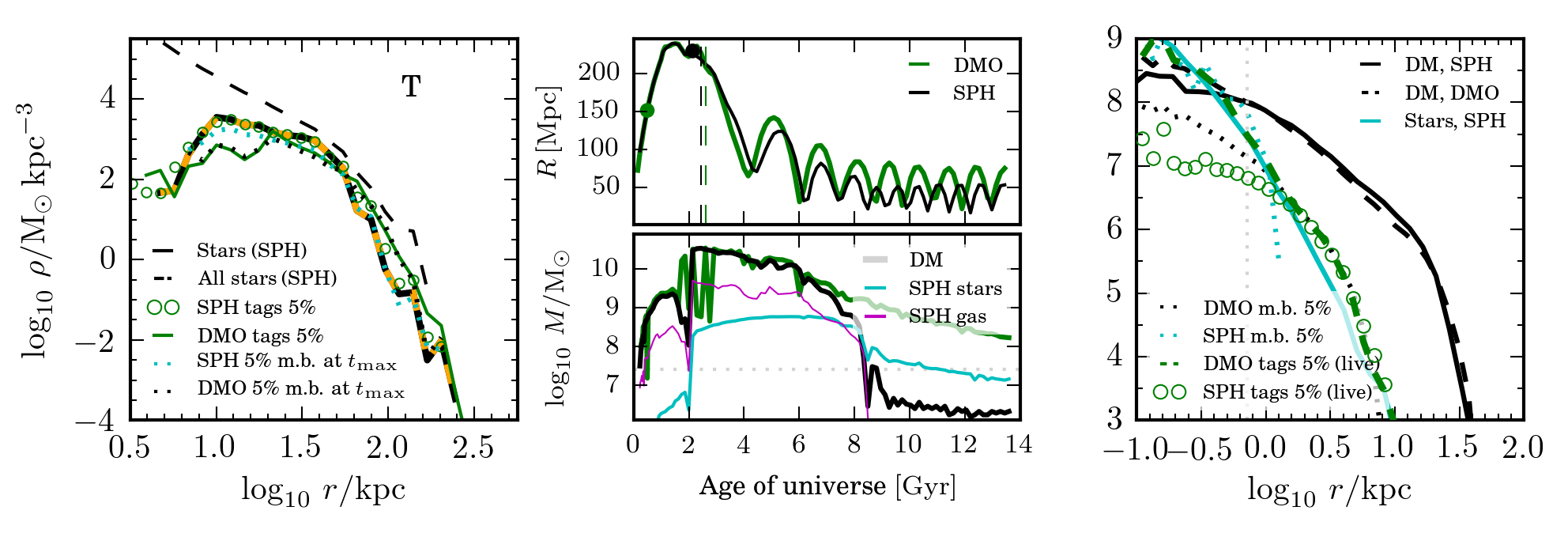}
  \includegraphics[width=150mm, trim=0.0cm 0cm 0.0cm 0cm,clip=True]{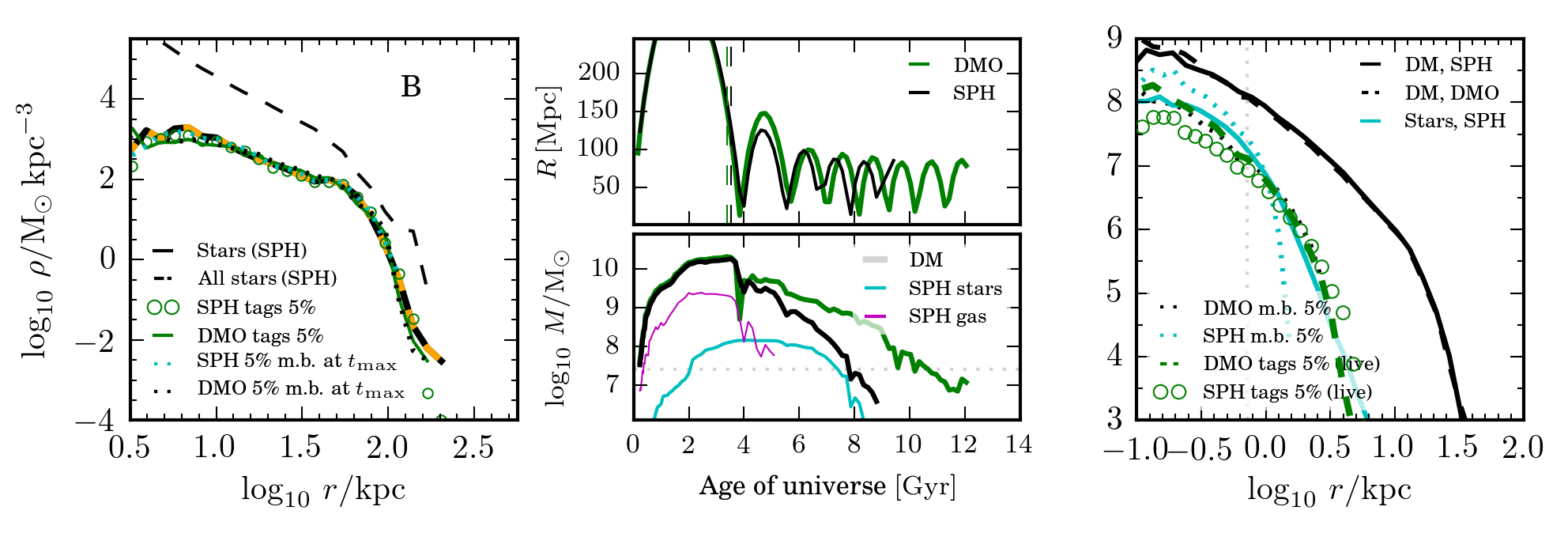}
  \includegraphics[width=150mm, trim=0.0cm 0cm 0.0cm 0cm,clip=True]{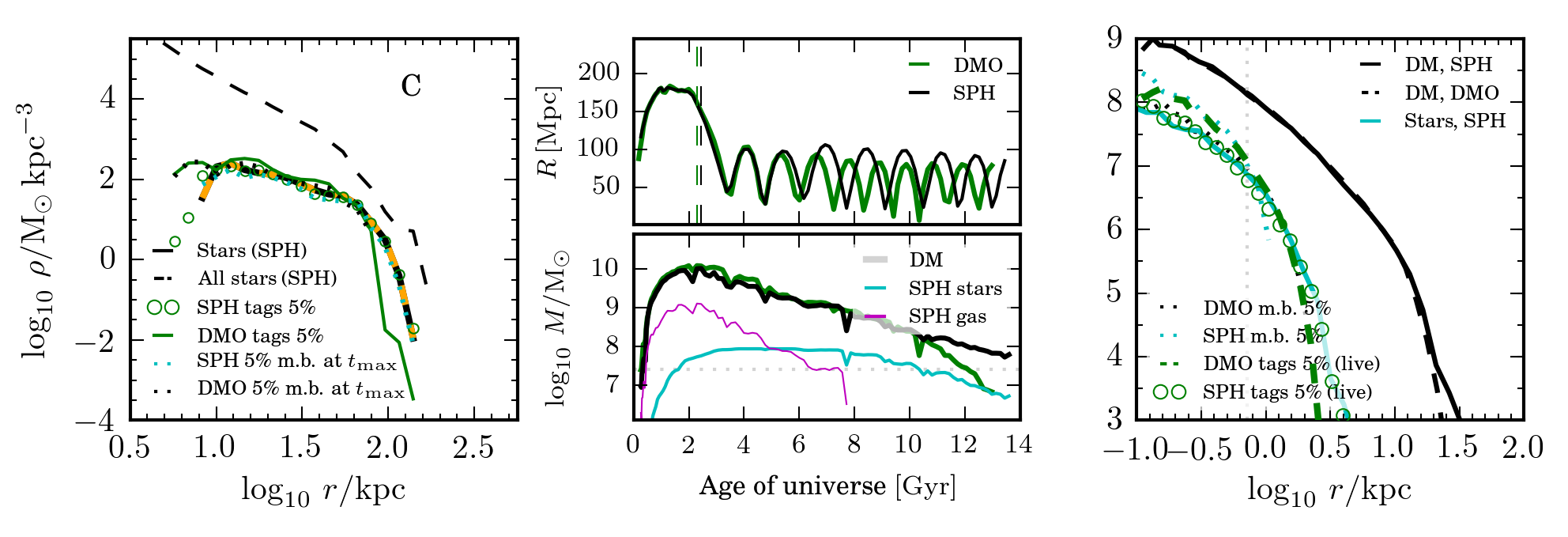}
  \includegraphics[width=150mm, trim=0.0cm 0cm 0.0cm 0cm,clip=True]{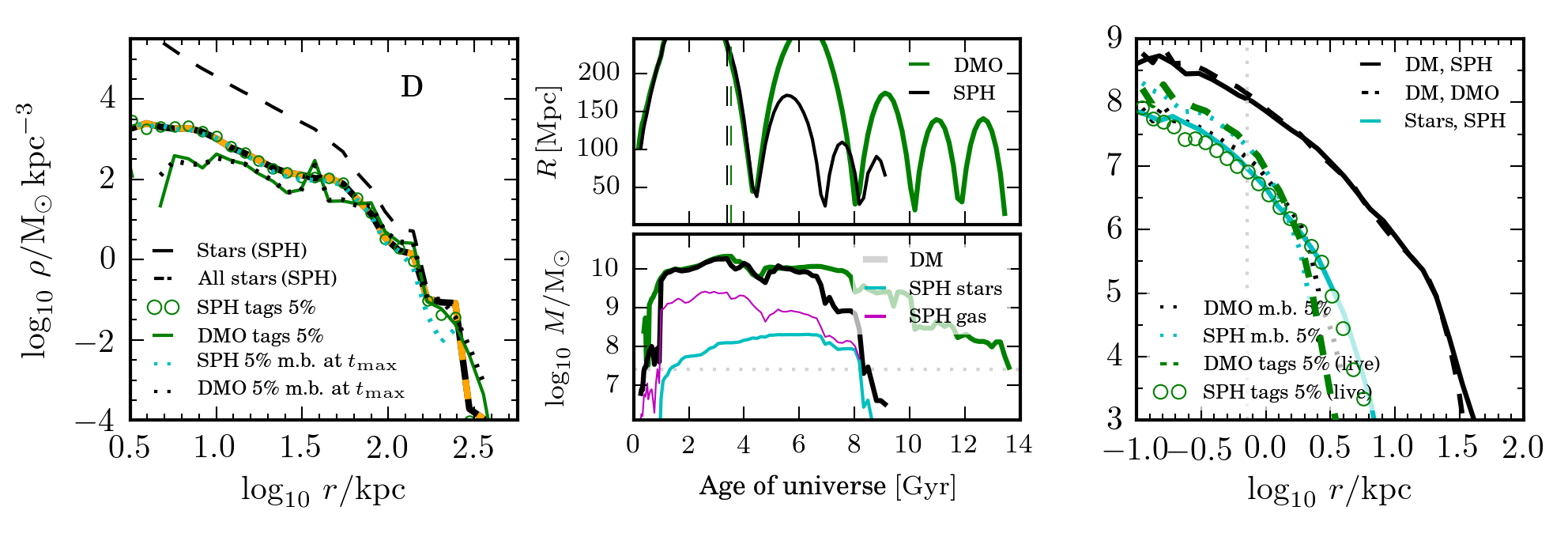}
  
  \caption{Further examples of individual stellar halo progenitors, comparing
  AqC-SPH and AqC-DM simulations. Panels show: \textit{left}, density profile
  of debris at the present day; \textit{centre top}, orbit of the progenitor
  (dashed line at infall time); \textit{centre bottom}, evolution of its
  gaseous, stellar and dark mass (dotted line at the mass of 10 particles);
  \textit{right}  density profile of the satellite halo at the time of maximum
  mass ($t_{\mathrm{max}}\approx t_{\mathrm{infall}}$).  Dotted black and blue
  lines in left- and right-hand panels correspond to a selection of the most
  bound 5~per~cent of the DM at the time of infall; dashed green line and circles in
  the right-hand panel correspond to the actual distribution of `tagged' DM in
  AqC-DM and AqC-SPH respectively.}

  \label{fig:examples_BCDE}
\end{figure*}

\begin{figure*}
  \includegraphics[width=150mm, trim=0.0cm 0cm 0.0cm 0cm,clip=True]{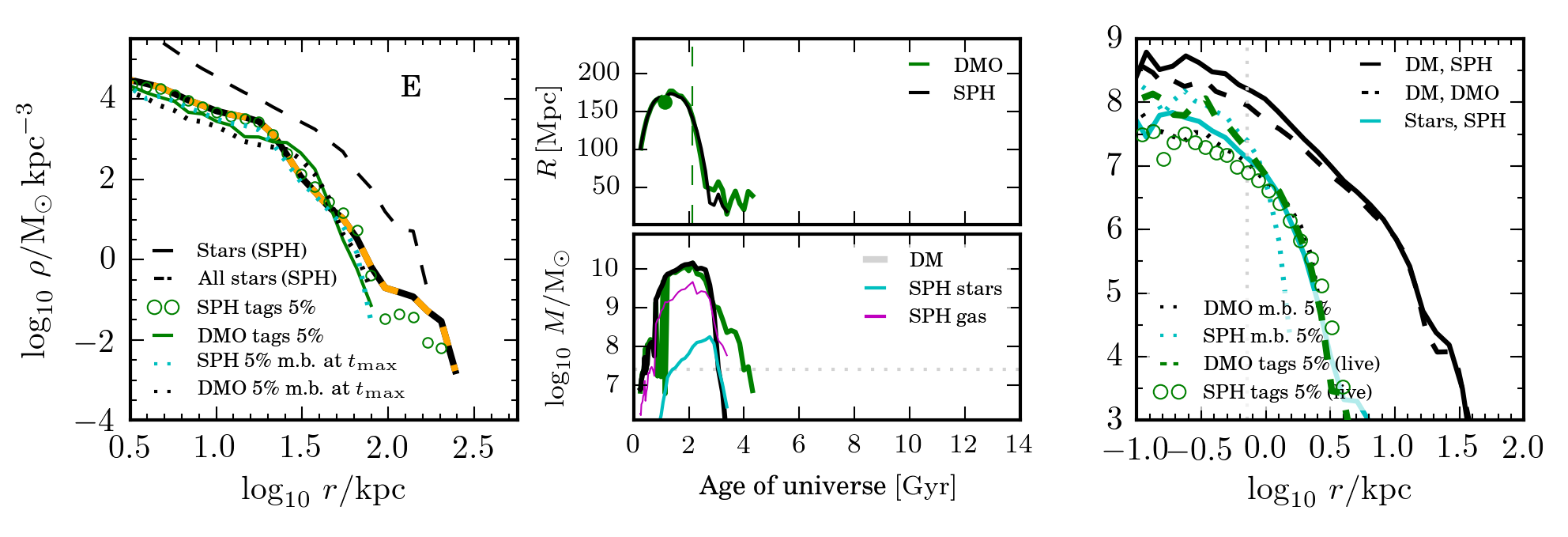}
  \includegraphics[width=150mm, trim=0.0cm 0cm 0.0cm 0cm,clip=True]{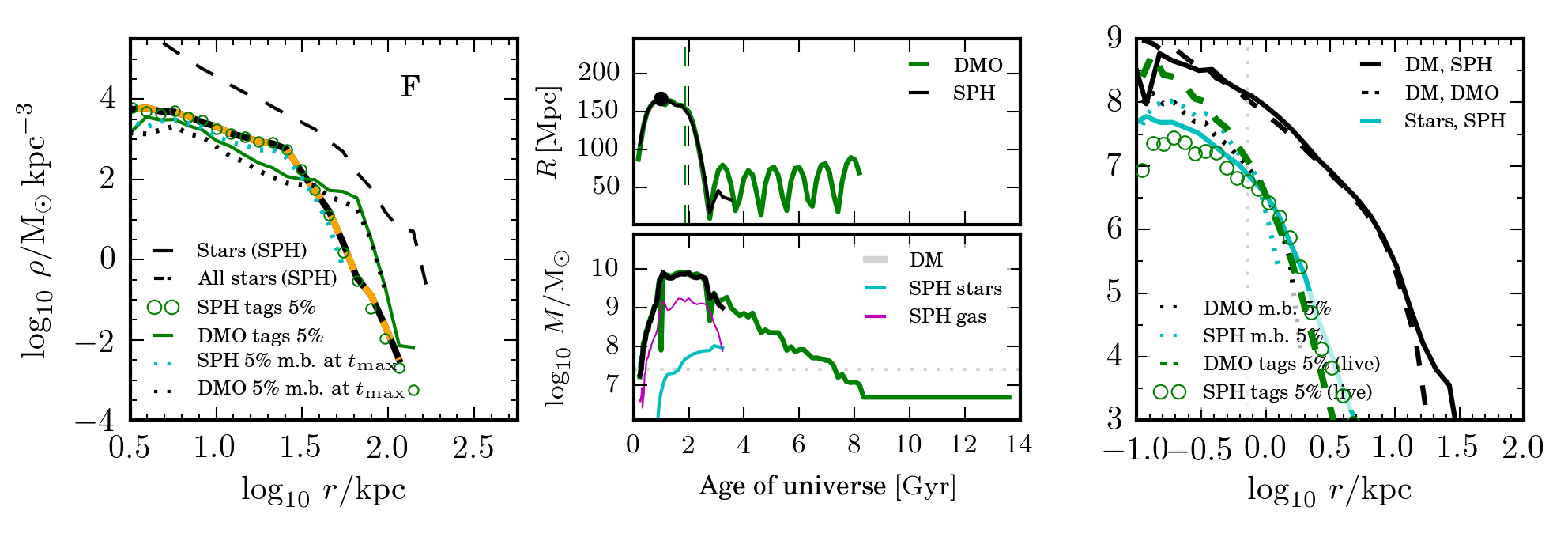}
  \includegraphics[width=150mm, trim=0.0cm 0cm 0.0cm 0cm,clip=True]{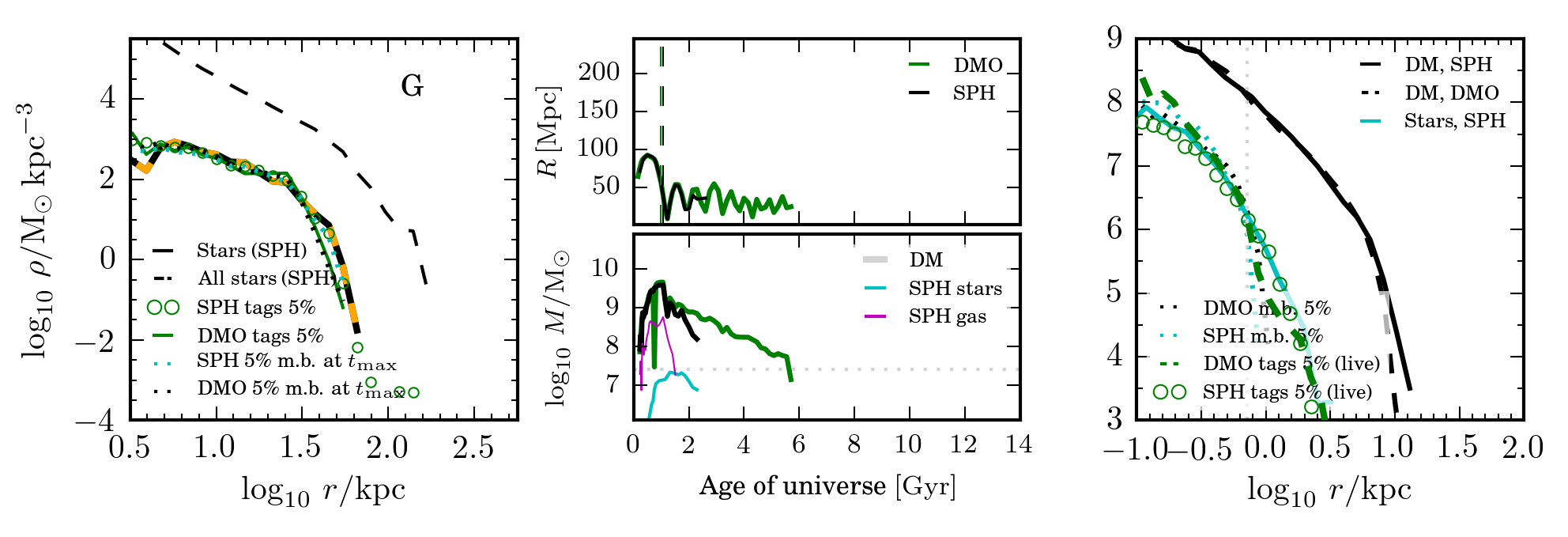}
  \includegraphics[width=150mm, trim=0.0cm 0cm 0.0cm 0cm,clip=True]{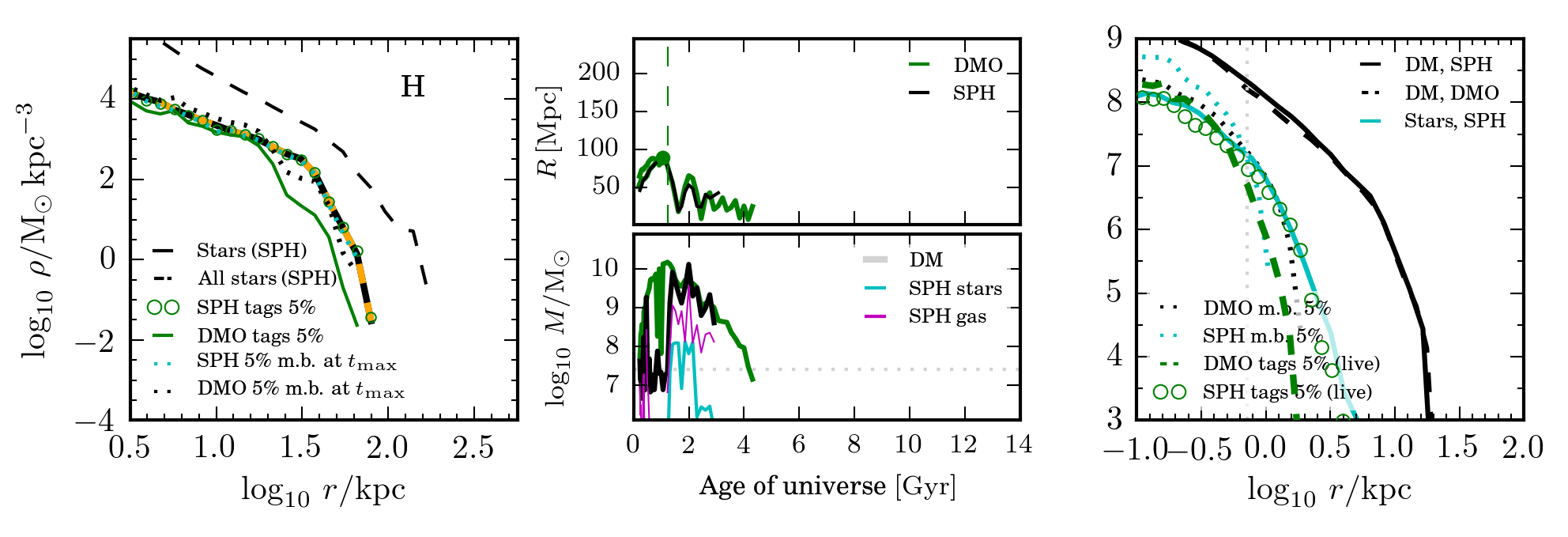}
  
  \caption{Continued.}

  \label{fig:examples_FGHI}
\end{figure*}

\section{Comparing SPH and \galform{} star formation models}
\label{appendix:galform_sfh}

\begin{figure}
  \includegraphics[width=84mm, trim=0.0cm 0cm 0.0cm 0cm,clip=True]{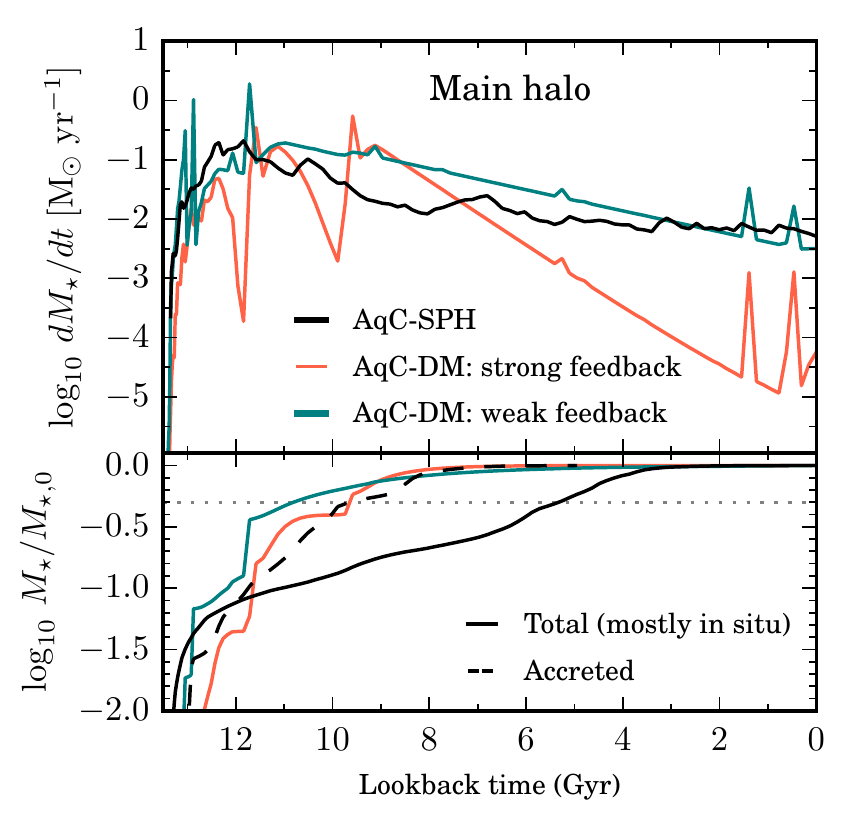}
  \includegraphics[width=84mm, trim=0.0cm 0cm 0.0cm 0cm,clip=True]{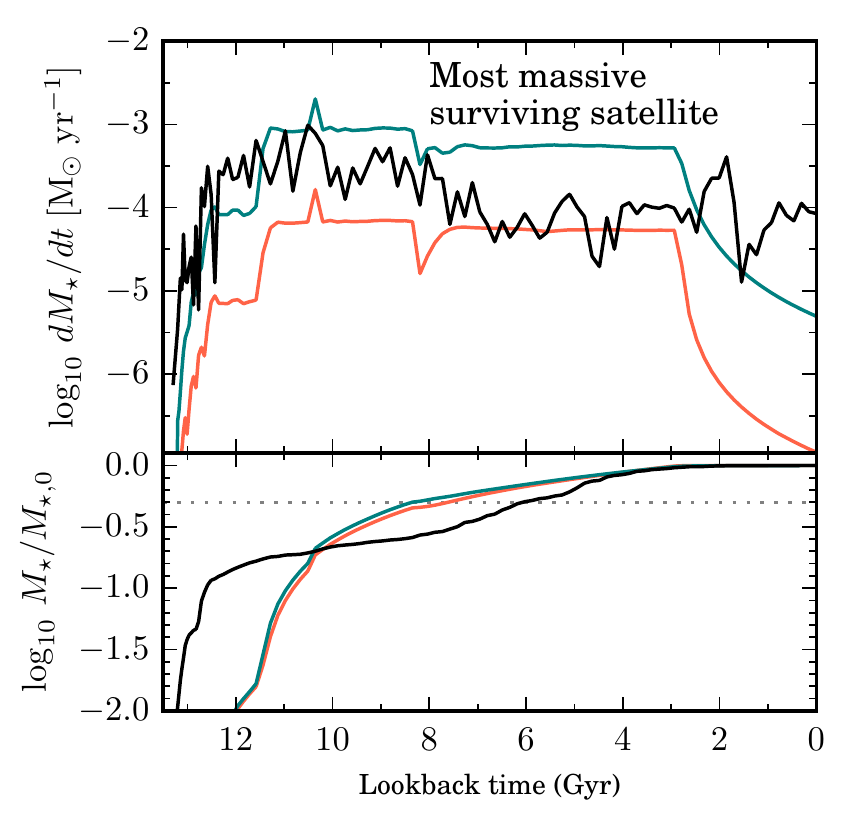}

  \caption{Top: \ins{} SFHs (upper panel) and fraction of $z=0$ \ins{} stellar
  mass in place (lower panels) as a function of lookback time for the MW
  analogue in our \sphsim{} simulation (black).  These are compared with SFHs
  from two semi-analytic \galform{} models, with relatively `strong' (red) and
  `weak' (blue) feedback (see text), applied to the \sphsim{} and \dmsim{}
  simulations. The black dashed line in the lower panel is the collective SFH
  of star particles in the accreted stellar halo at $z=0$ in \sphsim{}.
  Bottom: the same, for the most massive satellite of the MW analogue at
  $z=0$.}

\label{fig:sfh_compare}
\end{figure}

In Section~\ref{sec:galform_tagging} we presented particle tagging results for
the \dmsim{} simulation based on SFHs predicted by our semi-analytic code,
\galform{}. Since there is little reason to expect that \galform{} with our
fiducial choice of parameters will predict SFHs similar to those of \sphsim{},
we examined two choices of \galform{} parameters, which correspond to
relatively `weaker' and `stronger' feedback.  The weak feedback choice results
in a total stellar mass for the MW analogue and a density profile for its
accreted stellar halo more comparable to that of \sphsim{}, whereas the strong
feedback case corresponds to the parameter set used by C10\footnote{The set of
parameters we call `strong feedback', essentially the model of
\citet{Bower:2006aa}, was used by C10 because it can match a number of
observational constraints from the wider galaxy population when applied to a
representative cosmological volume; the weak feedback variant, all other
parameters being held fixed, would substantially overpredict the number of
galaxies with $L\lesssim L_{\star}$.}.  Interestingly, the good agreement in
Fig.~\ref{fig:c4_main_galform} suggests the SFHs produced by \galform{} for the
entire merger tree of \dmsim{} are not substantially different from those
predicted by the full hydrodynamic calculation.  

Fig. \ref{fig:sfh_compare} confirms this similarity by comparing the
\galform{} SFHs for the main halo in our simulation and its most massive
\textit{surviving} satellite against those from \sphsim{}. For both the weak and
strong feedback variants, \galform{} predicts more bursts of star formation in
the main halo at redshifts $z>2$, which leads to more rapid growth of the MW 
analogue relative to the \sphsim{} calculation.  Conversely, the amplitude
of the \galform{} SFR at later times is relatively low, but approximately
constant, as in \sphsim{}. The lower panels of Fig.~\ref{fig:sfh_compare} show
that, in the main halo, the rapid rate of \ins{} mass growth at high redshift
in the \galform{} models more closely resembles that of the stars in \sphsim{}
that are eventually accreted (black dashed line) than the stars formed \ins{}
in \sphsim{}. 

The agreement of SFR predictions for the most massive satellite appears
slightly better, although the total stellar mass is underpredicted in our
strong feedback variant. This overall increase in stellar mass is responsible
for the somewhat better agreement between the weak feedback variant and
\sphsim{} with regard to the amplitude of the stellar halo density profile,
shown in Fig.~\ref{fig:c4_main_galform}. In \sphsim{}, this satellite halo
forms the bulk of its stellar mass at very high redshift, but still
considerably earlier than \galform{} predicts.  These differences in the rate
of growth of the stellar mass are important, because they determine the
characteristic scale of the DM halo at the time of tagging and hence
the initial scale radius of the \ins{} population.  All else being equal,
earlier peak star formation will lead to the stellar tags being assigned to
more tightly-bound DM particles and hence more compact density profiles at
$z=0$.  For these two cases, where we see that the time-scale of \ins{} star
formation is significantly longer in \sphsim{}, that effect most likely
contributes to the more concentrated \ins{} components seen for \dmsim{} in
Fig.~\ref{fig:c4_main_galform}.

\section{Satellite sizes and surface brightness profiles}
\label{appendix:sizes}

\begin{figure*}
\includegraphics[width=84mm, trim=0.0cm 0cm 0.0cm 0cm,clip=True]{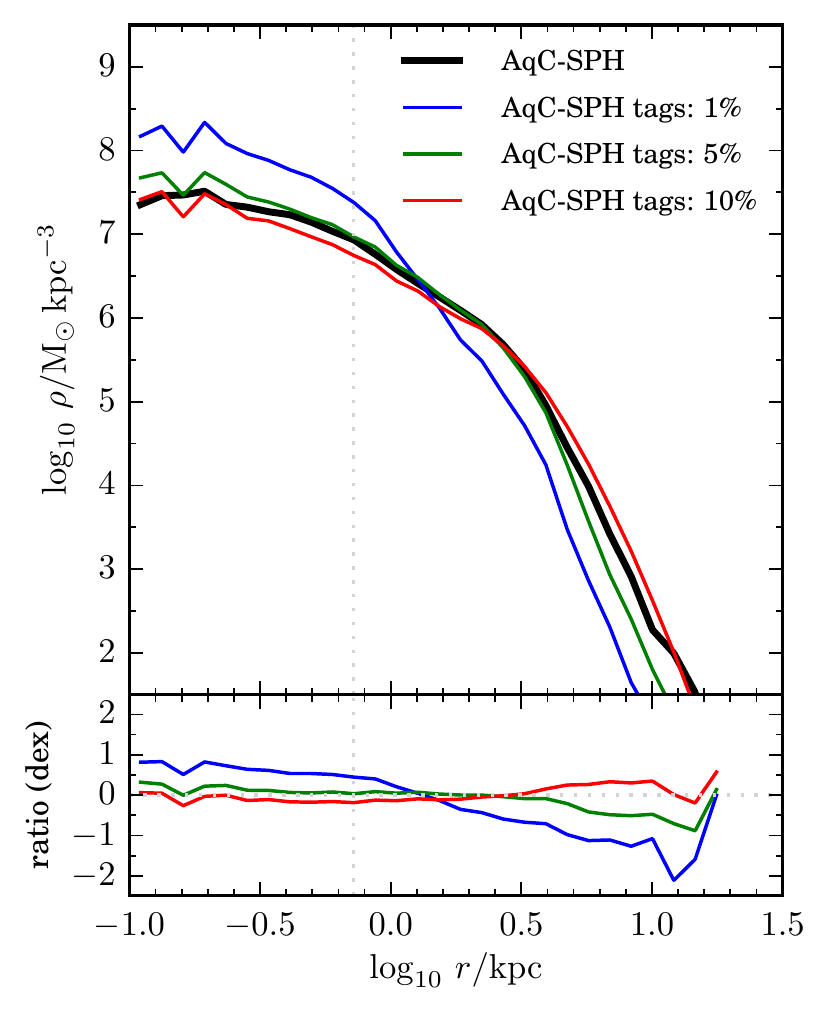}
\includegraphics[width=84mm, trim=0.0cm 0cm 0.0cm 0cm,clip=True]{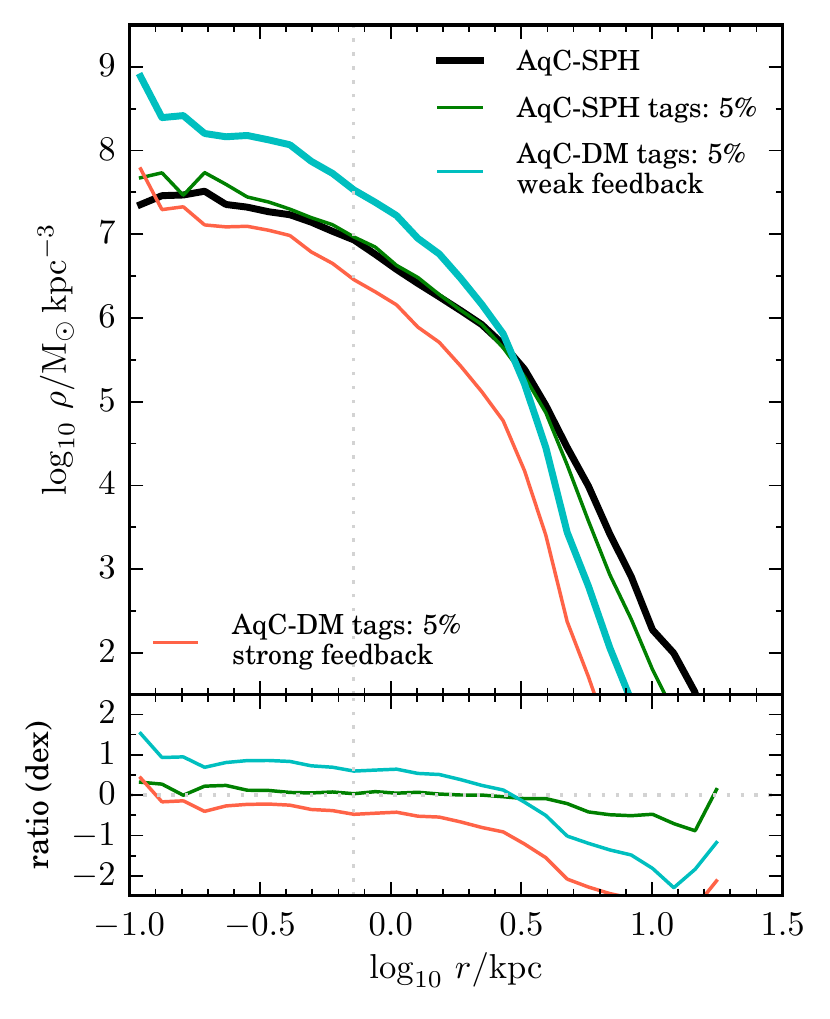}

\caption{Left: stellar mass density profiles as Fig.~\ref{fig:c4_main_std} for
the most massive satellite halo of our MW analogue at $z=0$. The lower
panel shows the ratio of stellar mass density in tagged particles to that in SPH
star particles. Right: a similar comparison with the results of tagging
($\fmb=5$ per cent) based on a \galform{} model in \dmsim{} (cyan), as in 
Fig.~\ref{fig:c4_main_galform}.}

\label{fig:c4_sat_comparison}
\end{figure*}

\begin{figure}
\includegraphics[width=84mm, trim=0.0cm 0cm 0.0cm 0cm,clip=True]{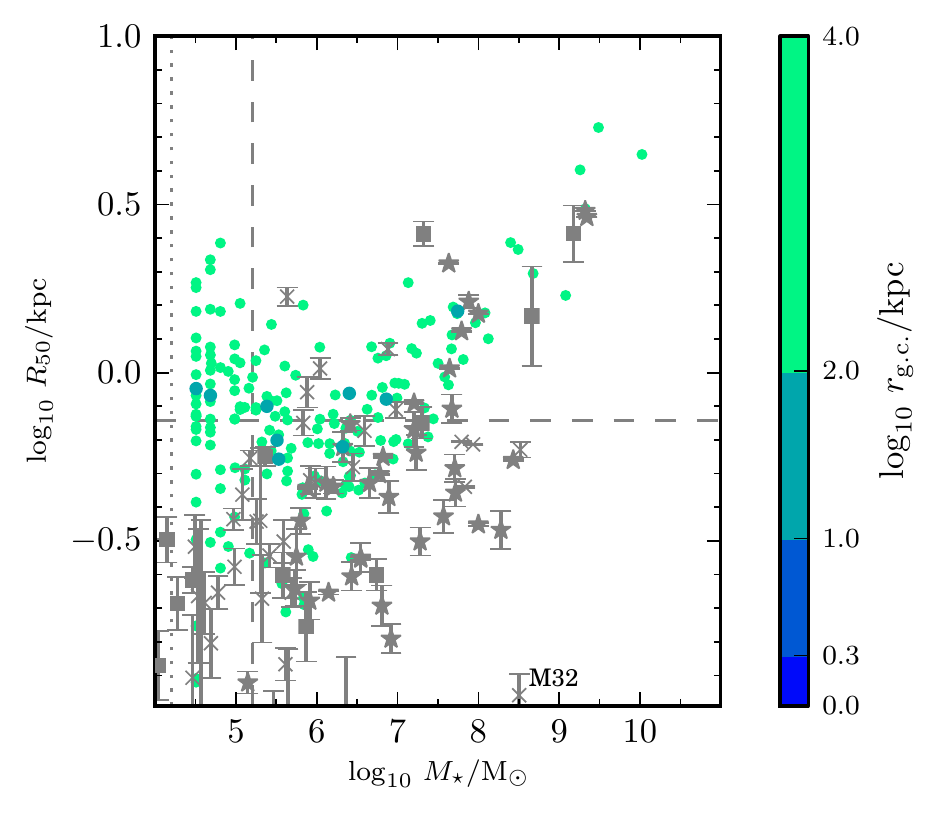}

\caption{Half mass radii of individual satellites in our SPH simulation as a
function of their stellar mass (points), colour-coded by their distance from the
central galaxy. Grey symbols with error bars show corresponding data for the
dwarf galaxies around the MW (squares), around M31 (crosses) and in the
Local Group (stars) for which both mass and size measurements are available in
the compilation of \citet{McConnachie:2012aa}. In cases where the error in half-mass
radius is unknown (17 of 95 objects), an error of $\pm50$ per cent is assumed.
The vertical and horizontal dashed lines indicate the limits in mass and size
below which the finite resolution of the simulation renders these results
unreliable. The horizontal dashed line shows the force softening scale of our
simulation.  At low mass, the apparent discretization is due to the
approximately quantized mass of individual star particles; vertical lines
correspond to the mass of 1 (dotted) and 10 (dashed) star particles.}

\label{fig:c4_hmr_data}
\end{figure}

\begin{figure*}
  \includegraphics[width=82mm, trim=0.0cm 0cm 0.0cm 0cm,clip=True]{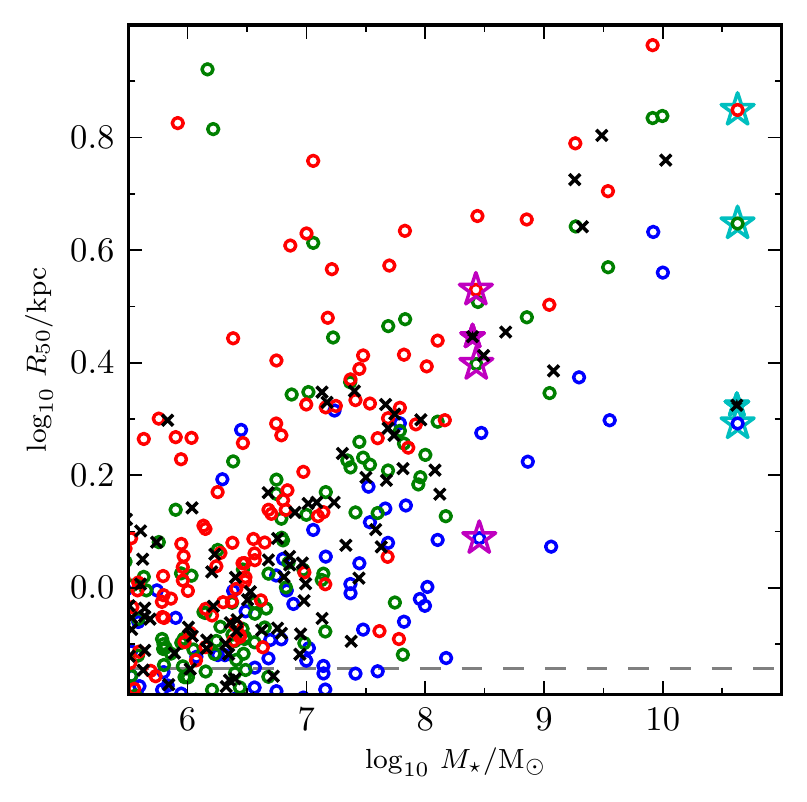}
  \includegraphics[width=82mm, trim=0.0cm 0cm 0.0cm 0cm,clip=True]{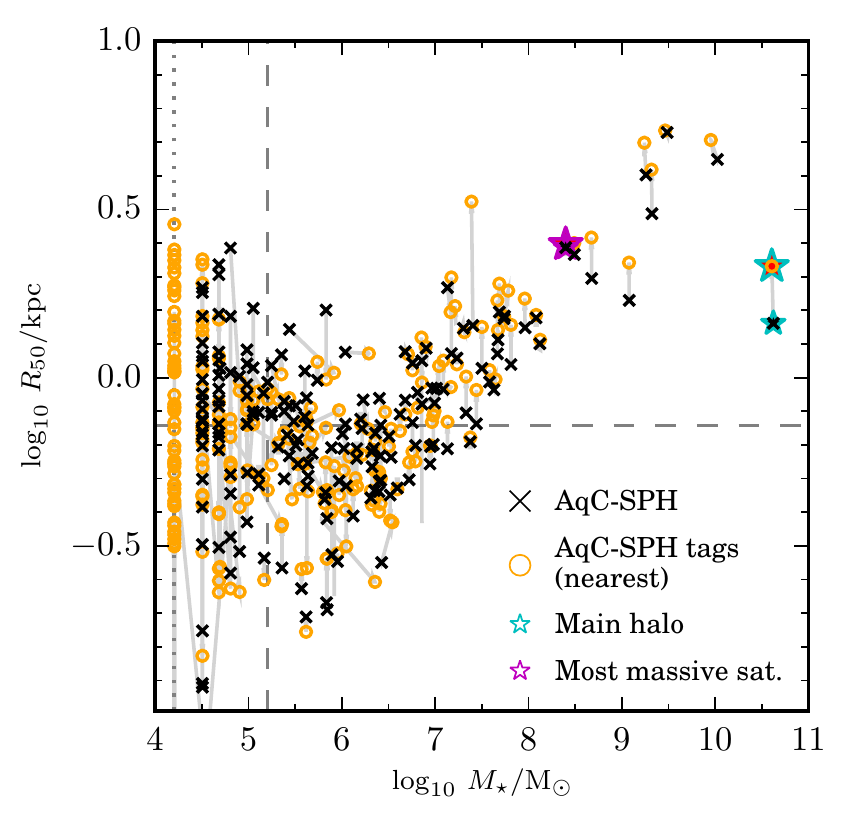}

  \caption{Satellite size--mass relations. Black crosses show the SPH results.
  Left: comparison against \stings{} fixed-fraction tagging with different
  $\fmb$: red 10~per~cent, green 5~per~cent and blue 1~per~cent. The set of four
  points (three tagging and one SPH) corresponding to the main halo are marked
  with cyan stars; likewise, points for the most massive satellite are marked
  with magenta stars. Right: comparison with `nearest energy neighbour' tagging
  scheme (orange). The central galaxy and its largest satellite are indicated
  as in the left-hand panel.  We have excluded subhaloes with low-resolution
  particles. Solid grey lines link the \sphsim{} point for each galaxy to its
  corresponding tagged particle realization. Broken grey lines mark the
  softening scale (horizontal dashed) and the mass of 1 and 10 gas particles
  (vertical dotted and dashed respectively).} 
  \label{fig:c4_hmr} 
\end{figure*}

The main text focuses on the MW analogue halo in our \sphsim{} and
\dmsim{} simulations. Fig.~\ref{fig:c4_sat_comparison} shows comparisons
between stars and tagged particles for the most massive surviving satellite
subhalo, analogous to Figs.~\ref{fig:c4_main_galform} and
\ref{fig:c4_main_std}. This is relevant because the size-mass relation of
surviving satellites provides an important constraint on the choice of $\fmb$.
Although the precise choice of $\fmb$ does not have a strong effect on the
distribution of stripped stars (Fig.~\ref{fig:c4_main_std}), it directly
determines the scale length of the \ins{} component in satellites that are not
strongly perturbed by tidal forces (as noted by C10 and elaborated on by C13).

The connection between $\fmb$ and the sizes of surviving satellites is shown in
the left-hand panel of Fig.~\ref{fig:c4_sat_comparison}.  Clearly, for this
particular satellite, $\fmb\sim5$~per cent is close to the optimal choice.  In
the right-hand panel of Fig.~\ref{fig:c4_sat_comparison}, we contrast this
result with tagging of \dmsim{} based on \galform{} ($\fmb\sim5$).  The impact
of the different SFHs in our two \galform{} variants is
clear, changing the amplitude of the profile (i.e. the total mass of stars in
the satellite) by an order of magnitude. Neither variant reproduces the
\sphsim{} profile very closely, suggesting that baryonic effects on the
potential and/or the orbital evolution of this satellite may differ
significantly between \sphsim{} and \dmsim{}. 

C10 and C13 calibrated $\fmb$ according to the median relation between
half-mass radius, $R_{50}$, and stellar mass,  $M_{\star}$.
Fig.~\ref{fig:c4_hmr_data} shows this relationship for galaxies in \sphsim{}.
Half-mass radii are measured from the centre of the potential of each subhalo
as reported by \subfind{} \citep{Springel:2001aa}. This figure also shows data
from galaxies in the Local Group to demonstrate that the distribution of sizes
for well-resolved galaxies in \sphsim{} are consistent with those of real dwarf
galaxies of similar mass. In very low mass haloes harbouring the smallest
galaxies in \sphsim{}, the potential is artificially cored by the gravitational
softening, which artificially inflates the sizes.

C10 found that $\fmb=1$~per cent resulted in a size--mass relation in
reasonable agreement with the Local Group observations shown in
Fig.~\ref{fig:c4_hmr_data}, although this also corresponded to the lower limit
of convergence with numerical resolution for their simulations. For a lower
resolution simulation of a much larger volume and using a different
semi-analytic model, C13 found values in the range 2--5 per cent best matched
the field galaxy size--mass relation for late-type galaxies. Given the
agreement between \sphsim{} and these observations, it is not surprising that a
similar value of $\fmb\sim5$~per cent best reproduces the results of the
subgrid star formation model in \sphsim{}.

The size--mass relations predicted by the tagging model variants discussed in
this paper are shown in Fig.~\ref{fig:c4_hmr}. The left-hand panel shows the
relations that result from tagging \sphsim{} with $\fmb = 1,5$ and $10$ per
cent. The main halo and the most massive surviving satellite discussed in the
previous subsection are highlighted by star symbols.

The right-hand panel of Fig.~\ref{fig:c4_hmr} shows similar results for the idealized `nearest
neighbour' tagging scheme discussed in Section~\ref{sec:nearest}. Grey lines
link the two representations of each satellite in this figure. For the majority
of well-resolved satellites (those in the upper right quadrant marked by dashed
grey lines) the tagged particle representation has systematically smaller
half-mass radius compared to its \sphsim{} star particle counterpart, although
the difference is small ($\lesssim 0.1$~dex). Conversely, tagged particles
bound to the main halo (rightmost point, highlighted) have a larger half-mass
radius than the corresponding star particles. The two points representing the
most massive satellite are highlighted with pink stars in
Fig.~\ref{fig:c4_hmr}; both figures show that the distribution of tagged
particles is slightly more compact than that of the corresponding star
particles. 

\section{Convergence}

\begin{figure}
  \includegraphics[width=84mm, trim=0.0cm 0cm 0.0cm 0cm,clip=True]{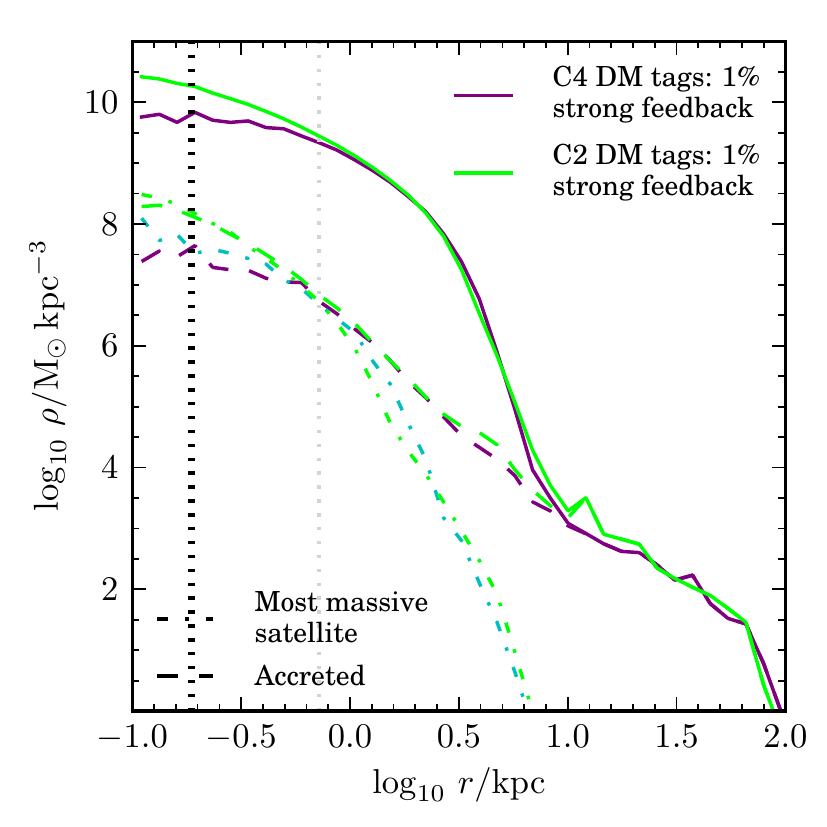}

  \caption{Convergence of the density profiles shown in
  Fig.~\ref{fig:c4_main_galform}. The green line results from application of
  the `strong feedback' model to a collisionless simulation with identical
  initial conditions to \dmsim{} and particle mass reduced by a factor of $20$.
  The dot--dashed line shows the density profile of the most massive satellite.}

\label{fig:c4_main_restest}
\end{figure}

Fig.~\ref{fig:c4_main_galform} shows the density profiles of the main halo and
its most massive satellite that result from application of the \galform{} models
discussed in this paper to a higher resolution version of our collisionless
simulation (\dmsim{}), with a particle mass $\sim20\times$ lower. This is the
resolution level used by C10. Fig.~\ref{fig:c4_main_galform} demonstrates that
the particle tagging results we are concerned with here have converged at the
resolution limit of \dmsim{}.

\label{lastpage}
\end{document}